\documentclass[12pt,preprint]{aastex}

\def\be{\begin{equation}}
\def\ee{\end{equation}}

\def\ergs{{\rm\,erg\,s^{-1}}}
\newcommand{\msun}{{M}_{\sun}}

\catcode`\@=11 
\def\@versim#1#2{\vcenter{\offinterlineskip
        \ialign{$\m@th#1\hfil##\hfil$\crcr#2\crcr\sim\crcr } }}
\def\mpy{M_{\sun} \ {\rm yr^{-1}}}

\shorttitle{****}
\shortauthors{****}
\begin{document}

\title{Revisiting the ``Fundamental Plane'' of Black Hole Activity at Extremely Low Luminosities}

\author{Feng Yuan\altaffilmark{1}, Zhaolong Yu\altaffilmark{1,2}, and
Luis C. Ho\altaffilmark{3}}
\altaffiltext{1}{Key Laboratory for Research in Galaxies and Cosmology,
Shanghai Astronomical Observatory, Chinese Academy of Sciences,
80 Nandan Road, Shanghai 200030, China; fyuan@shao.ac.cn}
\altaffiltext{2}{Graduate
School of the Chinese Academy of Sciences, Beijing 100039, China}
\altaffiltext{3}{The Observatories of the Carnegie Institution of Washington,
813 Santa Barbara Street, Pasadena, CA 91101, USA}

\begin{abstract}

We investigate the origin of the X-ray emission in low-luminosity
AGNs (LLAGNs). Yuan \& Cui (2005) predicted that the X-ray emission
should originate from jets rather than from an advection-dominated
accretion flow (ADAF) when the X-ray luminosity $L_{\rm X}$ of the
source is below a critical value of $L_{\rm X,crit} \approx
10^{-6}L_{\rm Edd}$. This prediction implies that the X-ray spectrum
in such sources should be fitted by jets rather than ADAFs.
Furthermore, below $L_{\rm X,crit}$ the correlation between radio
($L_{\rm R}$) and X-ray ($L_{\rm X}$) luminosities and the black
hole mass ($M$)---the so-called fundamental plane of black hole
activity---should deviate from the general correlation obtained by
Merloni, Heinz \& Di Matteo (2003) and become steeper. The Merloni
et al. correlation is described by ${\rm log}L_{\rm R} =0.6{\rm
log}L_{\rm X}+0.78{\rm log}M+7.33$, while the predicted correlation
is ${\rm log}L_{\rm R}=1.23{\rm log}L_{\rm X} +0.25{\rm
log}M-13.45$. We collect data from the literature to check the
validity of these two expectations. We find that among the 16 LLAGNs
with good X-ray and radio spectra, 13 are consistent with the Yuan
\& Cui prediction. For the 22 LLAGNs with $L_{\rm X} < L_{\rm
X,crit}$, the fundamental plane correlation is described by ${\rm
log}L_{\rm R}=1.22{\rm log}L_{\rm X}+0.23{\rm log}M-12.46
$, also in
excellent agreement with the prediction.
\end{abstract}


\section{Introduction}

Understanding the radiative quiescence of massive black holes has
important implications for accretion physics, fueling and feedback
mechanisms, and black hole growth over cosmic time (Ho 2004;
Pellegrini 2005). A viable physical model for the low luminosity of
most nearby active galactic nuclei (AGNs) is widely believed to be
due to the radiative inefficiency of advection-dominated accretion
flows (ADAFs; Narayan \& Yi 1994; 1995; see Ho 2008 for a review on
observational evidence on this point) present in these systems.
Moreover, the X-ray radiation of most low-luminosity AGNs (LLAGNs)
is believed to originate from ADAFs (see Yuan 2007 for a review).
LLAGNs are generally radio-loud (Ho 2008). The radio radiation of
AGNs is under-predicted by ADAFs and is widely accepted to be from
jets (e.g., Ho 2008 and references therein).  Merloni et al. (2003;
see also Falcke et al.  2004) find a so-called fundamental plane of
black hole activity, which is a correlation between the black hole
mass and radio and X-ray luminosities: \be {\rm log}L_{\rm
R}=(0.6^{+0.11}_{-0.11}){\rm log}L_{\rm X}+(0.78^{+0.11}_{-0.09})
{\rm log}M+7.33^{+4.05}_{-4.07}. \label{eq:merloni03} \ee The unit
of luminosity is $\ergs$, and the black hole mass $M$ is in units of
$M_\odot$. This finding is based on the radio-X-ray correlation of
some black hole X-ray binaries found by Gallo et al. (2003) and
Corbel et al. (2003) (but see Xue \& Cui 2007). The scatter of the
correlation, however, is quite large.

Yuan \& Cui (2005; hereinafter YC05) successfully interpret this
correlation based on an ADAF-jet model (to be described in \S3). In
this model, the X-ray radiation comes from the Comptonization
process in ADAFs while the radio originates from synchrotron
emission in jets. Obviously a key parameter here is the fraction of
the accretion material that goes into the jet, $\dot{M}_{\rm
jet}/\dot{M}$. YC05 find that to explain the correlation requires
$\dot{M}_{\rm jet}/\dot{M} \approx$  constant for $\delta=0.01$ and
$\dot{M}_{\rm jet}/\dot{M} \propto \dot{M}^{-0.5}$ for $\delta=0.5$.
Here, $\delta$ is a parameter in the ADAF model that describes the
fraction of turbulent dissipation that directly heats the electrons
(see Fig. 5 in YC05). Most of the sources in the sample of Merloni
et al. (2003) are relatively luminous. By extrapolating the above
$\dot{M}_{\rm jet}$ vs. $\dot{M}$ relation to lower accretion rates,
YC05 predict that the X-ray emission of the system should originate
from jets rather than ADAFs when the X-ray luminosity in the 2--10
keV band, $L_{\rm 2-10keV}$, is lower than a critical value $L_{\rm
X,crit}$: \be {\rm log}\left(\frac{L_{\rm X,crit}}{L_{\rm
Edd}}\right) =-5.356-0.17{\rm log}\left(\frac{M}{\msun}\right).
\label{eq:critlum} \ee The physical reason is as follows. The X-ray
emission from the ADAF is due to thermal Comptonization of the
synchrotron photons in the ADAF, which is roughly proportional to
$\dot{M}^2$. The synchrotron emission from the jet is proportional
to the density or the mass flux of the jet, $\dot{M}_{\rm jet}$. For
$\dot{M}_{\rm jet}\propto \dot{M}$ or $\dot{M}_{\rm
jet}\propto\dot{M}^{0.5}$, as $\dot{M}$ decreases, the synchrotron
emission from the jet will catch up with the Comptonization emission
from the ADAF and finally dominate the X-ray emission below a
certain $\dot{M}$, which corresponds to $L_{\rm X,crit}$ defined in
eq. (\ref{eq:critlum}) above. YC05 predict that below $L_{\rm
X,crit}$ the correlation should correspondingly steepen into the
following form: \be {\rm log}L_{\rm R}=1.23{\rm log}L_{\rm
X}+0.25{\rm log}M-13.45. \label{eq:yc05correlation} \ee

We would like to emphasize two points here. First, the work of YC05
is based on two assumptions: (1) that jets always exist in LLAGNs
and (2) that the physics of the jet does not depend on the
luminosity of the sources (it remains the same even when the
luminosity becomes extremely low). Second, all the above-mentioned
correlations, and the critical luminosity defined by eq.
(\ref{eq:critlum}) hold only in statistical sense. Given the large
scatter in the normalization of the primary fundamental plane
relation (Merloni et al. 2003), it is possible that any individual
source may not follow the correlation well, or that its critical
luminosity may differ significantly from that described by eq.
(\ref{eq:critlum}).  For the sake of concreteness, throughout this
paper we neglect this complication, although we should keep this in
mind when we quantitatively assess observations. Note, however, that
the slope of eq. (\ref{eq:yc05correlation}) is not affected because
it is determined only by jet physics.

Different from the above scenario, some people propose that
the X-ray emission {\it always} comes from jets, irrelevant to the
luminosities of the sources. Markoff et al. (2001) model
the broadband spectrum of the relatively luminous hard state of
XTE J1118+480 using a jet model. In this work, both the radio and
X-ray emission come from the jet. Markoff et al. (2003) subsequently
interpret the Merloni et al. correlation described by eq. (1) using this
jet model. Note that, different from YC05, no break in the correlation is
expected at lower luminosities in this model. In addition, the correlation
spectral index between radio and X-ray is calculated to be $\sim 0.6$,
i.e., consistent with eq. (1). This is again different from YC05
(see also Heinz 2004), where when both radio and X-ray emission originate
from a jet, the correlation spectral index is calculated
to be $\sim 1.23$ (see eq. 3). As Heinz (2004) points out, and confirmed
later in YC05, the discrepancy arises because Markoff et al. do not
take into account the cooling effect in the electron's steady energy
distribution.

The validity of the prediction of YC05 can be checked in two ways.
The first is by modeling the X-ray spectra of sources with X-ray
luminosities $L_{\rm X}<L_{\rm X,crit}$. In the usual ${\rm log}(\nu
L_{\nu})$ vs. ${\rm log}\nu$ spectral plot, if the X-ray spectrum
originates from a jet, it forms a straight line (i.e., a power-law
spectrum); if, on the other hand, it originates from an ADAF, the
spectrum should be curved.  Wu et al. (2007) model eight FR I
sources using a coupled jet-ADAF model. The X-ray luminosities of
these sources are usually low, $L_{\rm X}/L_{\rm Edd}\approx
10^{-4}-10^{-8}$. They find that the X-ray emission in the brightest
source ($L_{\rm X}/L_{\rm Edd} \approx 1.8\times 10^{-4}$) is from
an ADAF. The results for the four sources with moderate luminosities
are complicated, with two being dominated by an ADAF, one by jets,
and the other fitted by a combination of the two. The X-ray emission
of the three least luminous sources is dominated by a jet. Modeling
data from deep {\it XMM-Newton}\ observations of two black hole
transients in quiescence, Pszota et al. (2008) find that the X-ray
spectra are of power-law shape. These spectra are expected if the
jet dominates the X-ray emission, but they deviate significantly
from the expected spectrum of an ADAF at very low $\dot{M}$.

The second way to check the YC05 prediction is to look at the radio-X-ray
correlation. In this context, Pelligrini et al. (2007) present a
multiwavelength study of the quiescent black hole in the elliptical galaxy
NGC~821, combining radio data from the VLA and X-ray data from {\it Chandra}.
This source is very dim, with
$L_{\rm 2-10keV} \approx 10^{-8}L_{\rm Edd} \ll L_{\rm X,crit}\approx 2\times
10^{-7}L_{\rm Edd}$.  They find that the source follows the correlation
described by eq. (3) much more closely than that of eq. (2). Wrobel et al.
(2008) analyze similar data for two additional quiescent black holes in the
elliptical galaxies NGC~4621 and NGC~4697. For both sources, they find that
the radio fluxes predicted by eq. (3) are in relatively good agreement (within
a factor of 3) compared to the observed values.  Wang et al. (2006) and
Li et al. (2008) reinvestigate the fundamental plane relation using a
significantly larger sample of AGNs selected from the Sloan Digital Sky
Survey.  They find that the correlation slope for radio-loud AGNs is $1.39$,
steeper than the slope of 0.85 for the radio-quiet AGNs.  Since in radio-loud
AGNs essentially all of the radio emission and a significant fraction of the
X-ray emission come from jets, whereas in radio-quiet sources the X-ray
emission is dominated by the accretion flow, the steeper slope found for
radio-loud AGNs is qualitatively consistent with the prediction of YC05.
However, obviously more direct data for sources with $L_{\rm X} < L_{\rm X,crit}$
would be highly desirable.

In the present paper, we collect data from the literature to check the
prediction of YC05. We use a new sample of 22 LLAGNs with well-measured black
hole masses and radio and X-ray luminosities (Table 1) to reexamine the
fundamental plane correlation in the regime $L_{\rm X} < L_{\rm X,crit}$
(\S3).  In a complementary approach, we then model the spectral energy
distributions of another set of sources (Table 2) using an accretion-jet
model to investigate the origin of the X-ray emission (\S4). Section 5
summarizes our main conclusions.

\section{The ``Fundamental Plane'' of Black Hole Activity at Extremely Low
Luminosity}

\begin{figure} \epsscale{0.45} \plotone{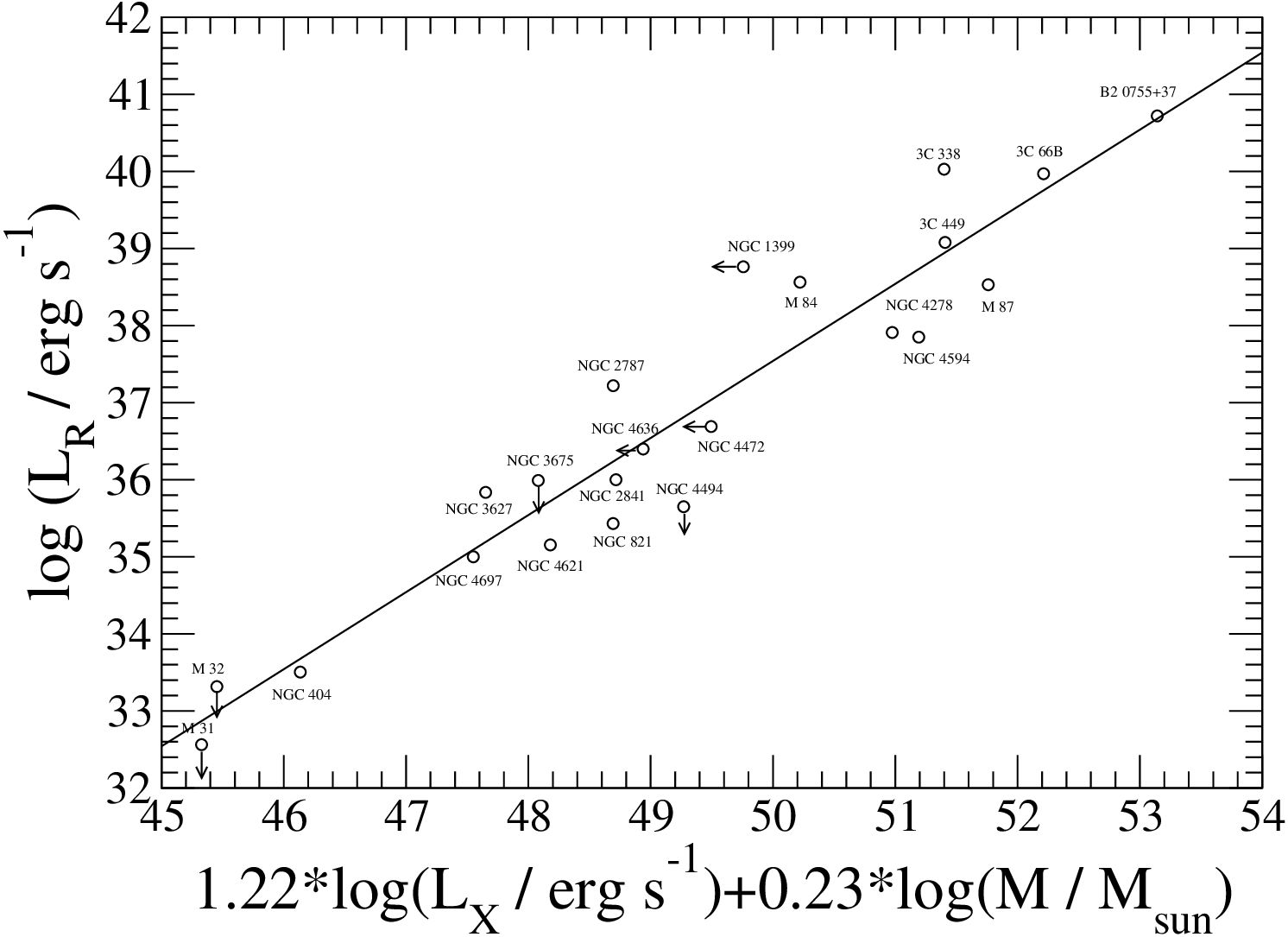}\hspace{1cm}
\epsscale{0.45}\plotone{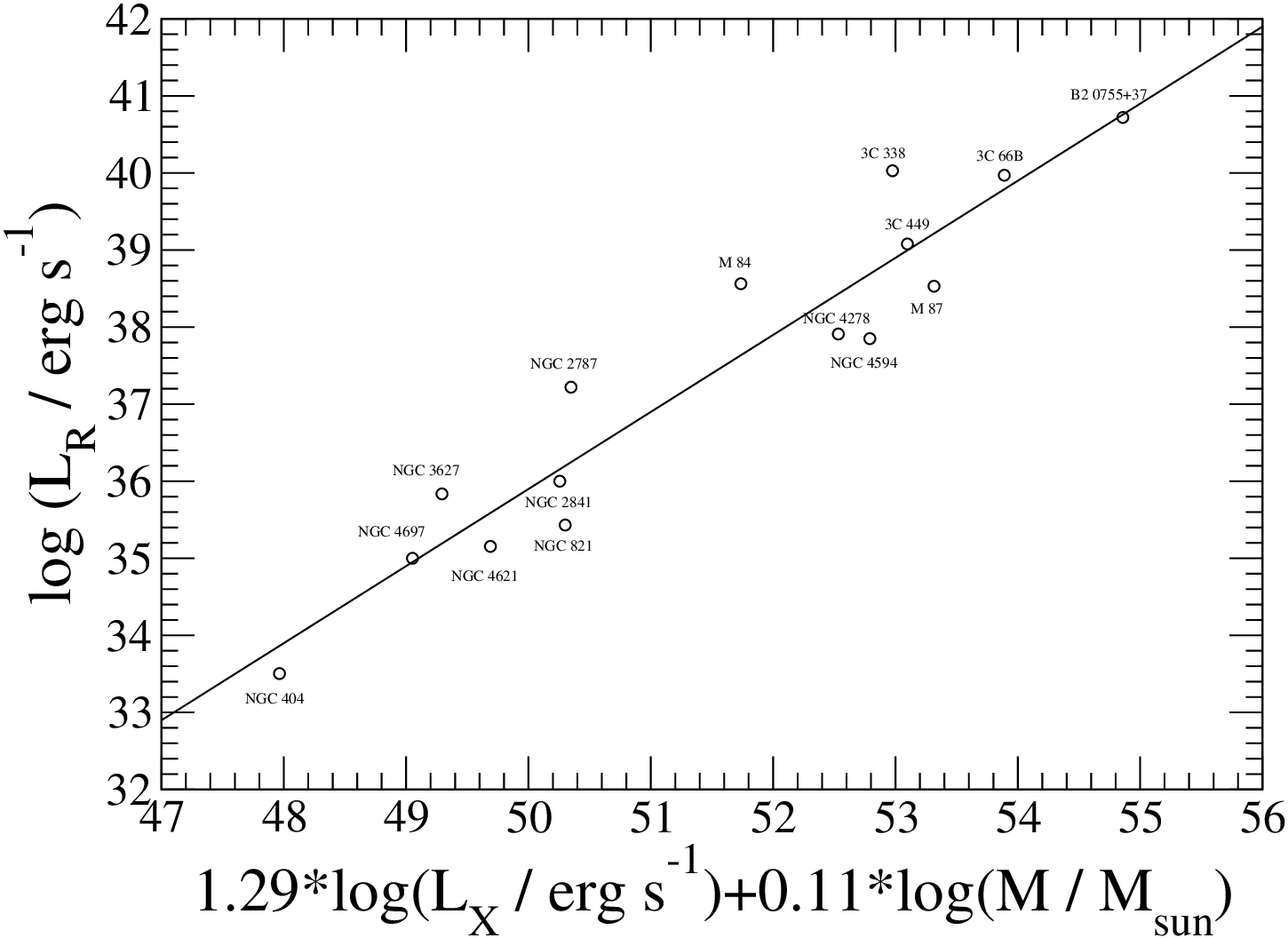} \vspace{0.5in}
\caption{The correlation among the mass and the radio and X-ray luminosities
for black hole sources with $L_{\rm X} <L_{\rm X,crit}$, as defined by eq. (1).
{\em Left:} The correlation for all 22 sources listed in Table 1.
Note we only have upper limits for the X-ray or radio luminosities for
seven sources. The solid line shows the best fit described by eq. (4).
{\em Right:} The correlation for the 15 sources listed in Table 1 without the
above-mentioned seven sources.  The solid line shows the best fit described by
eq. (5), which is in excellent agreement with the prediction of YC05.}
\label{correlation}
\end{figure}

To examine whether LLAGNs follow the correlations described by
eq. (1) or eq. (3), we collect black hole masses, radio (5 GHz) luminosities,
and X-ray (2--10 keV) luminosities for sources with
$L_{\rm X}<L_{\rm X,crit}$.  We find 22 sources that meet this criterion
(Table 1), as
shown by the left plot of Figure \ref{correlation}.  We do not include
Sgr A*, even though its luminosity satisfies our criterion. This is because
current observations indicate that there is no jet in this source, and thus
it violates the assumption of YC05.
Within our sample, seven sources (namely
M~31, M~32, NGC~1399, NGC~3675, NGC~4472, NGC~4494, and NGC~4636) only have
upper limits for their radio or X-ray luminosities.
For three sources (NGC~404, NGC~4621, and NGC~4697), we only
have radio luminosities at frequencies other at 5 GHz.
For these objects we deduce the luminosity than 5 GHz
either by extrapolating the observed radio spectrum (NGC~404)
 or, when no radio spectrum is available (NGC4621 and NGC~4697), by assuming
the radio spectrum predicted by our jet model.

We use the approach adopted by Merloni et al. (2003) and Wang et al.
(2006) to analyze the correlation between radio and X-ray
luminosities and the black hole mass. The solid line in the left
plot of Figure \ref{correlation} shows the best fit to all the 22
sources. It is described by \be {\rm log}L_{\rm R}=1.22(\pm
0.02){\rm log}L_{\rm X}+0.23(\pm0.03){\rm log} {M}-12.46.
\label{observedcorre} \ee If we exclude the seven sources with upper
limits, the correlation for the remaining 15 sources is shown by the
solid line in the right plot of Figure \ref{correlation}. It is
described by \be {\rm log}L_{\rm R}=1.29(\pm 0.03){\rm log}L_{\rm
X}+0.11(\pm0.04){\rm log} {M}-14.1. \ee The above results clearly
indicate that for sources with $L_{\rm X}<L_{\rm X,crit}$ the
correlation index between $L_{\rm X}$ and $L_{\rm R}$ is in much
better agreement with the prediction of YC05 than Merloni et al.
(2003).

\section{The ADAF-Jet Model}

We briefly describe the ADAF-jet model here. The readers can refer
to Yuan et al. (2005) for additional details.

\subsection{The ADAF Model}

The accretion flow consists of two parts. Within a ``transition'' radius
 $R_{\rm out}$ the flow is described by an ADAF (sometimes also called
a ``radiatively inefficient accretion flow''; Narayan \& Yi 1994,
1995). Outside of $R_{\rm out}$, the accretion flow is described by
a standard thin disk. Both observations and theoretical studies
indicate that the value of this radius is a function of accretion
rate (Liu et al. 1999; Meyer et al. 2000; Yuan \& Narayan 2004).
Numerical simulations (Stone et al. 1999; Hawley \& Balbus 2002;
Igumenshchev et al. 2003) and analytical work (Narayan \& Yi 1994;
Blandford \& Begelman 1999; Narayan et al. 2000; Quataert \&
Gruzinov 2000) on ADAFs indicate that only a fraction of the gas
available at large radius actually accretes onto the black hole. The
rest of the gas is either ejected from the flow or is prevented from
being accreted by convective motions. We therefore parameterize the
accretion rate with a parameter $s$, defined such that
$\dot{M}=\dot{M}_{\rm out}(R/R_{\rm out})^{s}$, where $\dot{M}_{\rm
out}$ is the accretion rate at the boundary $R_{\rm out}$. There is
obviously a degeneracy between $\dot{M}_{\rm out}$ and $s$ when the
accretion rate at the innermost region of the ADAF is concerned. We
calculate the global dynamical solution of the ADAF. Other
parameters include the viscosity parameter $\alpha$ and magnetic
parameter $\beta$ [defined as the ratio of the gas to the total
pressure (sum of gas and magnetic pressure) in the accretion flow,
$\beta=P_{\rm g}/P_{\rm tot}$], and $\delta$, the fraction of the
turbulent dissipation that directly heats the electrons. The
radiative processes we consider include synchrotron, bremsstrahlung,
and their Comptonization. After the ADAF structure is obtained, the
spectrum of the flow can be calculated in the same way as in
previous work (e.g., Yuan et al. 2003). Specifically, the X-ray
radiation comes from the thermal Comptonization process.

In our calculation we adopt typical values of
$\alpha=0.3$ and $\beta=0.9$, which are widely used in ADAF models.
Although these two values are still not well constrained and some uncertainties
still exist, their values do not significantly affect the X-ray spectrum.
This is partly because their effects are absorbed in $\dot{M}_{\rm out}$
and $\delta$, respectively. The main parameter left in the ADAF model
is therefore $\delta$. The value of the parameter $\delta$ (and also $s$)
are well constrained in the case of our Galactic center black hole,
Sgr A* (Yuan et al. 2003). Assuming that the physics of the
ADAFs is the same for various objects, we thus follow Yuan
et al. (2003) and adopt $\delta=0.5$ and $s=0.3$ as ``fiducial'' values.
But we also try other values if we cannot get a good fit to the observations.
In summary, the X-ray spectrum, including the normalization and the
spectral slope, which is what concern us mostly in the present
work, is mainly determined by the combination of the accretion rate
at the innermost region of the ADAF (which is determined by the combination of
$\dot{M}_{\rm out}$, $R_{\rm out}$, and $s$) and the electron temperature
(which is determined mainly by $\delta$). Roughly speaking, there is a
one-to-one correspondance between the normalization and the spectral slope,
and the photon index predicted by an ADAF in the 2--10 keV band ranges from
$\Gamma\la 1.5$ at the highest luminosities to $\Gamma \ga 2$ at the
lowest luminosities (see, e.g., Fig. 3a in Esin et al. 1997).

\subsection{The Jet Model}

The jet model adopted in the present paper is based on the internal
shock scenario, which has also been widely used in the study of
gamma-ray burst afterglows. A fraction of the material in the
accretion flow is assumed to be transferred into the vertical
direction to form a jet. The mass lost rate is set as $\dot{M}_{\rm
jet}$. The jet is assumed to have a conical geometry with half-open
angle $\phi$ and bulk Lorentz factor $\Gamma_{\rm jet}$. Throughout
this paper we generally assume $\phi=0.1$ and $\Gamma_{\rm jet}=10$
if there is no observational constraint on them. Internal shocks
occur as a result of the collision between shells with different
velocities. As a result, a small fraction  of the electrons in the
jet are accelerated into a power-law energy distribution with index
$p$. Shock acceleration theory typically gives $3>p>2$. For example,
for relativistic shocks in Bednarz \& Ostrowski (1998) and Kirk et al.
(2000) obtain $p\approx 2.2$. However, there is still some
uncertainty in our understanding of shock acceleration, and, more
generally, the acceleration mechanism of electrons in jets; for
instance, magnetic reconnection may be another relevant mechanism in
addition to shocks.  So we also try values of $p<2$. In this
context, we note that the modeling of BL Lac objects, where the
emission comes from predominantly from jets, sometimes also requires
$p<2$ (e.g., Ghisellini et al. 2002). We assume that the fraction of
accelerated electrons is $\xi_{\rm e}$ and fix $\xi_{\rm e}=1\%$ in
our calculations. The energy density of accelerated electrons and
amplified magnetic field in the shock front are described by two
free parameters $\epsilon_{\rm e}$ and $\epsilon_{\rm B}$. Medvedev
(2006) shows that they should roughly follow $\epsilon_{\rm e} \sim
\epsilon_{\rm B}^{1/2}$.

Only synchrotron emission is considered in the calculation of the jet
spectrum. Compton scattering is neglected. This approximation is suitable,
as shown in Wu et al. (2007), when when $\dot{M}_{\rm jet}$ is small,
as is the case for all sources in the present work.  This is because,
on the one hand, when $\dot{M}_{\rm jet}$ is small,
the ratio of the photon energy density to the
magnetic field energy density is very low. Thus, the power of synchrotron
self-Compton emission in the jet is several orders of magnitude less than
that of synchrotron emission. On the other hand, $p\approx 2$ implies
that a significant fraction of the synchrotron emission power lies in the
X-ray band of our interest. The
radio spectrum emitted by the jet is due to self-absorbed synchrotron emission
from different parts of the jet, and the spectrum is usually flat (i.e.,
$\alpha \approx0$, where $F_{\nu}\propto \nu^{\alpha}$).  The X-ray emission
comes from optically thin synchrotron emission from accelerated electrons in
the jet, and the slope is mainly determined by the value of $p$. Since the
radiative cooling timescale of these electrons is typically much shorter than
the dynamical timescale, the spectral index of the {\em steady} distribution
of electrons is $p+1$. So the spectral index of the X-ray spectrum is
$\alpha=-(p+1-1)/2$, or $\alpha \approx -1$ for $p \approx 2$. The
normalization of the spectrum is mainly determined by the values of
$\epsilon_{\rm e}$ and $\epsilon_{\rm B}$.

\subsection{The Truncated Thin Disk}

We do not consider the contribution to the spectral energy distribution from a
truncated standard thin disk outside of $R_{\rm out}$. The main reason is that
we want to focus on the origin of the X-ray emission. For this
purpose, only the jet and ADAF are the most relevant since the radiation of
the standard thin disk only contributes up to the optical and ultraviolet (UV)
bands.  For some sources in our sample, we do not have reliable data in the
optical or UV.  Good data at optical and UV, of course, can present additional
constraints on the model parameters, but mainly to the combination of $R_{\rm out}$
and $\dot{M}_{\rm out}$. As we describe before, there is a degeneracy between
these two parameters when we model the X-ray emission from an ADAF, and so
the constraint from the optical and UV data to the focus of our paper is very
limited.  For these reasons, we only consider the jet and ADAF components and
only attempt to fit the radio and X-ray spectra.

\section{Modeling Results for Individual Objects}

We fit the radio and X-ray spectra of 16 sources using the ADAF-jet
model. The results are shown in Figures 2 and 3, and the model
parameters are listed in Table 2. The accretion rate of the ADAF and
the mass loss rate of the jet are in unit of $\dot{M}_{\rm
Edd}\equiv 10L_{\rm Edd}/c^2$.  Among these sources, seven are
dominated by ADAFs (Fig. 2) and nine by jets (Fig. 3). Regarding
the origin of the X-ray emission, all but three of these 16 sources
are consistent with the prediction of YC05. We describe the
fitting results below.

\subsection{ADAF-Dominated Sources}

This type of sources includes IC~4296, NGC~315, NGC~1052,
NGC~4203, NGC~4261, NGC~4579, and NGC~6251 (Fig. 2). The X-ray spectra of these
sources can be fitted very well by ADAFs, while the contribution from jets
is negligible. From Table 2, we see that for all these seven sources
$L_{\rm 2-10keV} \gg L_{\rm X,crit}$.  This is consistent with the prediction
of YC05.  Below are notes to some sources.

{\bf IC~4296.} The 0.3--10 keV X-ray
spectrum of this source consists of two components,
a power law that dominates above 1 keV with photon index
$\Gamma=1.48^{+0.42}_{-0.34}$ and soft thermal emission with
$kT=0.56^{+0.03}_{-0.03}$ keV that dominates below 1 keV.
Since the thermal component likely comes from the host galaxy background,
we only adopt the hard power-law component.
The accretion rate required in our model is roughly
consistent with the Bondi accretion rate
derived by Pellegrini et al. (2003).

{\bf NGC~315.}
Worrall et al. (2007) obtained X-ray spectra of
both the jet and the nucleus. The power-law component of the nucleus
is described by $\Gamma=1.57\pm0.11$, much harder that that of the jet,
which is $\Gamma=2.2\pm0.2$. The X-ray luminosity $L_{\rm 2-10keV}=
6\times 10^{41}\ergs\approx 1.5\times 10^{-5}L_{\rm Edd}$, which
is much higher than the critical luminosity of $L_{\rm X,crit}
\approx 1.6\times 10^{-7}L_{\rm Edd}$. Worrall et al. (2007)
argue that the X-rays from
the jet must have a synchrotron origin, which implies that the power-law index
of the electrons is $p=2.4\pm0.4$. The significant difference between
the spectrum of the nucleus and the jet can be considered as possible evidence that
the origin of the X-ray emission of the nucleus is from the accretion
flow rather than from the jet. We see from the figure that this is
consistent with our conclusion.

{\bf NGC 4261.} From analysis of the {\it Chandra}\ image, the X-ray jet
extends to the nucleus.   Zezas et al. (2005) find that the X-ray spectrum can
be well fitted by three components, namely a thermal component and two
power laws with $\Gamma_1=1.54^{+0.71}_{-0.39}$ and
$\Gamma_2=2.25^{+0.52}_{-0.28}$. Since the soft X-rays are
dominated by the thermal component, we only adopt the data of the two
power-law components, with the harder component being the dominant one.
As shown in the figure, our modeling indicates that
the dominant harder component comes from the ADAF while the softer one comes
from the jet (Worrall \& Birkinshaw 1994; Zezas et al. 2005), consistent
with the prediction of YC05.
This interpretation is consistent with the much lower absorption
of the softer power-law component, $N_{\rm H}<3.7\times 10^{20}\,{\rm cm}^{-2}$.
Zezas et al. (2005) point out that this is also supported by
the 1~keV flux of this component, which is within a factor of a few of that
estimated on the basis of the nuclear 4.8 GHz radio flux,
using the X-ray to radio ratio of knot A in the western jet.

{\bf NGC~4579.} Quataert et al. (1999) have already modeled this source. They
find that the X-ray spectrum can be fitted by the ADAF very well. Our
calculation is in good agreement with theirs, as shown in the figure.

{\bf NGC 6251.} Both {\it Chandra}\ and {\it XMM-Newton}\ observed this source
(Evans et al. 2005).  The spectra obtained appear to be mildly discrepant,
given the formally quoted error bars.
{\it Chandra}\ gives $\Gamma=1.67\pm0.06$ while {\it XMM-Newton}\ gives a
steeper spectrum with  $\Gamma=1.88\pm0.01$.  The reason for the discrepancy
is either genuine spectral variability or possibly residual photon pileup
effects suffered by the {\it Chandra}\ data.  We therefore only fit the
{\it XMM-Newton}\ data.

\subsection{Jet-Dominated Sources}

This type of sources includes IC~1459, M~32, M~81, M~84, M~87, NGC~3998,
NGC~4594, NGC~4621,
and NGC~4697 (Fig. 3).  The X-ray spectra (and radio, of course) of these
sources can be fitted very well by jets, but not
by the ADAF model. To illustrate this point,
we show in the figure by the dot-dashed lines the emitted spectra from an ADAF.
For M~32, M~81, M~84, M~87, NGC~3998, and NGC~4594, the parameters of the ADAF
are adjusted so that the model can produce the ``correct'' X-ray flux.
For the other three sources, the parameters of the ADAF are adjusted
so that the model does not violate the radio data.
>From Table 2, we see that for all sources except IC~1459, M~81, and NGC~3998,
$L_{\rm 2-10keV} \ll L_{\rm X,crit}$.  Consistent with the prediction of
YC05, the X-ray emission is produced by jets. For the three outliers,
$L_{\rm 2-10keV}> L_{\rm X,crit}$. However, as we emphasize in \S1, there is a
large error bar in the value of $L_{\rm X,crit}$ described by eq.
(\ref{eq:critlum}). In addition, $L_{\rm 2-10keV}$ of these three
sources are also not much different from $L_{\rm X,crit}$ predicted by
eq. (\ref{eq:critlum}) (ref. Table 2). Given the ill-defined errors on the
actual measurements and the known large scatter in the normalization of the
fundamental plane relation (and hence in the predicted value of
$L_{\rm X,crit}$), the exact value of $L_{\rm X}$/$L_{\rm X,crit}$ for
any particular source should not be interpreted too literally.
Below are notes to some sources.

{\bf IC~1459.} The X-ray emission is well fitted by a jet, while the
contribution of the ADAF is constrained by the radio data to be at very low
level. Given that $L_{\rm 2-10keV}>L_{\rm X,crit}$, the origin of X-ray
emission is apparently inconsistent with the prediction of YC05, but, as
mentioned above, we should be wary about overinterpretation of individual
sources, especially in this case where $L_{\rm X}$/$L_{\rm X,crit}$ is
only marginally greater than unity. The value of $p$ is 1.9, formally
but not significantly smaller than 2.
Fabbiano et al. (2003) also fit the X-ray spectrum with a jet model, but with
a much larger $p=2.78$.
The discrepancy of the value of $p$ is because that they do not
consider the effect of radiative cooling on the energy distribution of
electrons. In another words, the value of $p$ adopted there is the spectral
index of the {\em steady} distribution
rather than the {\em injected} distribution.

{\bf M~32.} The X-ray spectrum is well fitted by a jet.  The dot-dashed line
shows our best fit by an ADAF; the fit is only marginally acceptable. However,
the required accretion rate at the Bondi radius is
$\sim 8\times 10^{-6}\mpy$, which is about 10 times larger than
the Bondi rate $\dot{M}_{\rm Bondi}\approx (3-10)\times 10^{-7}\mpy$
estimated by Ho et al. (2003) from {\it Chandra}\ observations.
This situation is very similar to the cases of M~31 and the quiescent
state of the black hole X-ray binary XTE J1118+480 (YC05).
In the case of M~31, Garcia et al. (2005) estimated the Bondi accretion rate
from {\it Chandra}\ observations to be $\dot{M}_{\rm Bondi} \approx 6\times
10^{-6}\dot{M}_{\rm Edd}$. However, the
X-ray luminosity produced by an ADAF with such an accretion rate would be
4 orders of magnitude lower than the observed $L_{\rm X}$. For XTE J1118+480,
from optical observations together with disk instability
theory for the outburst, the mass accretion rate is estimated to be
$\dot{M}\approx 10^{-6}\dot{M}_{\rm Edd}$ (McClintock et al. 2003). Again,
an ADAF with such an accretion rate would underpredict the X-ray luminosity by
nearly 4 orders of magnitude (YC05).
If the X-ray radiation is dominated by a jet, on the other hand,
this will not be a problem at all. Taking M~32 as an example,
the mass accretion rate at the innermost region of the ADAF,
say at $5R_{\rm S}$, is about
$\dot{M}_{\rm Bondi}(5R_{\rm S}/R_{\rm Bondi})^{0.3}\approx 6\times 10^{-7}
\dot{M}_{\rm Edd}$. So we only require
$10^{-8}/6\times 10^{-7}\approx 2\%$ of the accretion matter
transferred into the jet to produce the correct X-ray flux.
The physical reason is that the radiative efficiency
of the jet is much higher than that of an ADAF.

{\bf M~81.} We cannot model the X-ray emission using an ADAF
component alone. In contrast to our result, Quataert et al. (1999)
can fit the X-ray spectrum with an ADAF. The reason for the
discrepancy is that the mass of the black hole they adopt is $\sim
20$ times smaller than ours. On the other hand, the X-ray spectrum
is well fitted by the jet (see also Markoff et al. 2008), although
the value of $p$ in the jet model (1.8) is again formally smaller
than 2.

{\bf M~87.}
The spectra of the nucleus and jet knots are very similar, and the  X-ray flux
of the knots closest to the nucleus is high. Wilson \& Yang (2002) therefore
suggest that the X-ray emission of the nucleus actually comes from the jet
rather than the accretion flow.  Our detailed modeling confirms their
speculation.  As shown in the figure, the X-ray spectrum is well fitted by the
jet, but not by an ADAF. Di Matteo et al. (2003), on the other hand, were able
to fit the X-ray spectrum with an ADAF. This is because they do not consider
the outflow in their ADAF model.

{\bf NGC~3998.} We find that it is hard to model the X-ray emission
using an ADAF with outflow because the predicted spectrum of an ADAF
is too hard.
This conclusion
is the same as in Ptak et al. (2004). If we abandon the requirement of
significant outflow and direct electron heating (i.e, $\delta\ll 1$),
we can fit the X-ray spectrum well, as in Ptak et al. (2004). But this
kind of ADAF model is not favored from a theoretical point of view.
On the other hand, a jet alone can fit the X-ray spectrum well, although
we again require $p< 2$ (1.8).

{\bf NGC~4621.}
The X-ray luminosity of this source is extremely low.
This makes the source suitable to test
the prediction of YC05. As pointed out by
Wrobel et al. (2008), application of the correlation of
YC05 (eq. 3 above) predicts the observed radio flux to
be $\nu L_{\nu}(8.5~{\rm GHz})=1.5\times 10^{35}\ergs$, which is in
very good agreement with the observed value of $\nu L_{\nu}(8.5~{\rm GHz})
=3.3\times 10^{35}\ergs$.

{\bf NGC~4697.} This is another good source to test the prediction
of YC05 since the luminosity is again extremely low.
Consistent with YC05,
the X-ray spectrum is fitted very well by a jet.
Like NGC~4621,  the application of the correlation of
YC05 (eq. 3 above) predicts the observed radio flux to
be $\nu L_{\nu}(8.5~{\rm GHz})=3.5\times 10^{35}\ergs$, which agrees well
with the observed value of $\nu L_{\nu}(8.5~{\rm GHz})
=1.3\times 10^{35}\ergs$, as pointed out by Wrobel et al. (2008).

\section{Summary and Discussion}

We collect data for a sample of LLAGNs to investigate the origin
of their X-ray emission. YC05 predict that when the $2-10$ keV luminosity of
the system is smaller than a critical luminosity
$L_{\rm X,crit}$ (see eq. (\ref{eq:critlum}) for its definition), the
X-ray radiation will be dominated by the jet rather than by the ADAF. In this
case, YC05 predict that the correlation between the mass of the black hole
and the X-ray and radio luminosities will have a relation described by
eq. (\ref{eq:yc05correlation}), which is
steeper than the correlation found by Merloni et al. (2003;
eq. \ref{eq:merloni03}). In this paper we examine the validity
of this prediction. We assemble black hole masses, radio luminosities, and
X-ray luminosities for 22 sources from the literature
with $L_{\rm 2-10keV}<L_{\rm X,crit}$ to investigate their correlation.
We find that the correlation is best described by eq. (\ref{observedcorre}),
which is very close to eq. (3) but much steeper than eq. (\ref{eq:merloni03}).
We also use our ADAF-jet model to fit the radio and X-ray spectral data of 16
sources covering a wider range in $L_{\rm 2-10keV}/L_{\rm X,crit}$.
We find that 13 sources are consistent with the prediction of YC05.

Fender et al. (2003) argue for a ``jet-dominated'' quiescent state in
accreting black hole systems,
in the sense that the kinetic power of the jet is much
greater than the X-ray luminosity of the accretion flow when the X-ray
luminosity is below a certain critical value. The similarity between
this work and YC05 is that both emphasize the importance of the jet
when the system is very dim. However, these two works are also
intrinsically different, as discussed in YC05.
Fender et al. compare the unobservable quantity ``jet power'' with
the observable ``X-ray luminosity.'' They do not address the origin of the
X-ray radiation of the system.

There is a dichotomy in the observed properties of the X-ray nuclei
of FR I and FR II radio galaxies (Evans et al. 2006). The X-ray
spectrum of FR Is is usually unabsorbed, while that of FR IIs is
associated with high absorption and (narrow) iron line emission. In
addition, the average photon index of the FR Is presented in Evans
et al. (2006) is $\Gamma\approx 2$, while that of FR IIs is
$\Gamma\approx 1.7$. Evans et al. (2006) speculate that this
dichotomy arises from the relative contributions of the jet and
accretion-related emission, which depends on the total power of the
source. In less luminous FR I sources, the X-rays come from the jet
and are largely devoid of absorption from the torus.  By contrast,
in FR II sources the X-rays come from the accretion flow, which is
at least partly obscured by the torus. The model of YC05 presents a
natural theoretical explanation to the above-mentioned observations
and speculations. This is because FR I sources usually have
luminosities below $L_{\rm X,crit}$ while FR IIs are much more
luminous (see also Wu et al. 2007). In addition, from Figures 2 and
3, we see that the spectrum produced by an ADAF is usually harder
than that of a jet. The latter predicts $\Gamma=1+(p+1-1)/2\approx
2$ if $p\approx 2$.

Gallo et al. (2006) obtained radio and X-ray data for the quiescent state of
the black hole binary system A0620$-$00, which has
$L_{\rm X}\approx 10^{-8.5}L_{\rm Edd}\ll
L_{\rm X, crit}\approx 10^{-5.5}L_{\rm Edd}$.
They find that this source lies on the extrapolation
of the ``general'' radio-X-ray correlation, which is dominated by
the sources V404 Cyg and GX 339$-$4. However, we need to
be cautious when we conclude that the above result is not consistent with
the YC05 prediction. As pointed out by Wu et al. (2007)
and Corbel et al. (2008),
because of the large scatter in the ``general'' correlation,
it is not appropriate to connect
one data point from {\em one} source with the data of {\em other}
sources. Unfortunately, data for the more luminous hard state of A0620$-$00 are unavailable, which hampers us to reach any reliable conclusion.

Corbel et al. (2008) obtained simultaneous radio and X-ray data for both the
hard and quiescent states of V404 Cyg. They find that the radio-X-ray
correlation holds from the hard state down to its quiescent state, although
the correlation slope index is $\sim 0.5$, which is smaller than the general
value of $0.6-0.7$. However, V404 Cyg is the brightest quiescent state
black hole system known
(Tomsick et al. 2003). The X-ray luminosity Corbel et al. (2008) observed
is $L_{\rm X}\approx 8\times10^{32}\ergs
\approx 6\times 10^{-7}L_{\rm Edd}$. This is only 5 times smaller than the critical
luminosity calculated from eq. (1), $L_{\rm X,crit}\approx
3\times 10^{-6}L_{\rm Edd}$, and thus the deviation of the correlation from
eq. (2) is not expected to be large. This fact, combined with the
large uncertainty of the theoretical model, leads us to think that it is hard
to make a robust conclusion.

In summary, current investigations indicate that the YC05 prediction
holds for most sources, including AGNs (YC05; Wu et al. 2007; Pellegrini et al.
2007; Wrobel et al. 2008; this work) and black hole X-ray binaries (YC05;
Pszota et al. 2008). Since the correlation only holds in a statistical sense,
it is not surprising that we find some outliers.

We close by noting that our results have an interesting implication
for galaxy evolution.  Feedback from the central black hole is now
widely believed to play an important role in galaxy formation (e.g.,
Croton et al. 2006).  The kinetic power of the jet provides an
important source of energy feedback, and it would be valuable if
this parameter can be estimated from observations. If the black hole
system is in its ``quiescent'' phase, with $L_{\rm X}\la L_{\rm
X,crit}$, as many nearby galaxies seem to be, we can estimate the jet
power $P_{\rm jet}$ from $L_{\rm X}$ because the observed X-ray
luminosity comes from jets. What we need to know for this purpose
 is the radiative efficiency of jets, which is defined as
$\eta_{\rm jet}\equiv L_{\rm X}/P_{\rm jet}$.
Specifically, $L_{\rm X} \approx \delta^3L_{\rm X,int}$ and $P_{\rm jet}=\Gamma_{\rm jet}^2
\dot{M}_{\rm jet}c^2$. Here $L_{\rm X,int}$ is the intrinsic luminosity
emitted by the jet in its comoving frame, $\Gamma_{\rm jet}$
is the Lorentz factor of the bulk motion of the jet, and $\delta
\equiv [\Gamma_{\rm jet}(1-\beta {\rm cos}\theta)]^{-1}$ is the Doppler
factor of the jet. So we have
$\eta_{\rm jet}=\delta^3L_{\rm X,int}/\Gamma_{\rm jet}^2\dot{M}_{\rm jet}c^2$.
For $1/\Gamma_{\rm jet}<\theta<1$, $\delta\approx\Gamma_{\rm jet}\theta^2$,
and thus
$\eta_{\rm jet}=\Gamma_{\rm jet}\theta^6 L_{\rm int}/\dot{M}_{\rm jet}c^2$.
>From our modeling of the sources listed in Table 2, we have
\be
\eta_{\rm jet}\approx 10^{-5}\frac{\Gamma_{\rm jet}}{10}
\left(\frac{\theta}{60^{\circ}}\right)^6.
\ee
Thus, the ``intrinsic'' radiative efficiency of the jet
$\eta_0\equiv L_{\rm X,int}/\Gamma_{\rm jet}^2
\dot{M}_{\rm jet}c^2\approx 10^{-4}(\Gamma_{\rm jet}/10)^{-2}$. It is
interesting to note that this is consistent with the value obtained by Celotti
\& Fabian (1993) ---
$\eta_0 \approx 10^{-4.1\pm2.2}$ --- from modeling of the parsec-scale jets based
on VLBI observations.

\begin{acknowledgements}

This work was supported in part by the Natural Science Foundation of China
(grants 10773024, 10833002, 10821302, and 10825314), Bairen Program of
Chinese Academy of Sciences, and the National Basic Research Program of China
(973 Program 2009CB824800).  The research of L.C.H. was supported by
the Carnegie Institution of Washington.

\end{acknowledgements}

\clearpage

\begin{table*}
\caption{Sample with $L_{\rm 2-10keV}<L_{\rm X,crit}$ Used for the
Fundamental Plane Analysis}
\label{tab 1}
\begin{tabular}{lcccccccccc}
\tableline
\hline
Object  & log $M$ &Ref. & log $L_{\rm R}$ & Ref. & log $L_{\rm 2-10keV}$ & Ref.  \\
(1) & (2) & (3) &(4) & (5) & (6) & (7)  \\
\hline
3C~66B   &8.8 &16,39 &40.0 &16,40 &41.0 &16,39&\\
3C~338   &9.2 &1 &40.0 &1 &40.3 &1&\\
3C~449   &8.4 &16,39 &39.1 &16,41 &40.5 &16,39&\\
B2 0755+37 &8.9 &16,39 &40.7 &16,38 &41.8 &16,39&\\
M~31 & 7.5 & 12,32 & $<32.6$ &12,35& 35.7&13&\\
M~32 & 6.4 & 3,21 & $<33.3$ & 3& 36.0 &3&\\
M~84 & 9.2 & 1,2,20 & 38.6 & 1& 39.3 &1&\\
M~87&9.5 & 4& 38.5 &6,30& 40.5 &9,10&\\
NGC~404 & 5.3 &  4& 33.5 & 4,24& 36.7 &5&\\
NGC~821 & 7.9 &11,31 & 35.4 &11& 38.3&11&\\
NGC~1399 & 9.1&  19& $<38.7$ & 18& 39.0 & 17&\\
NGC~2787 & 7.6& 27& 37.2 & 37,43& 38.4 & 14,44 &\\
NGC~2841 & 8.4 &15,36& 36.0 &15,37 &38.3&14,15&\\
NGC~3627 & 7.3 &15,36& 35.8 &15,37  &37.6&14,15&\\
NGC~3675 & 7.1 & 36&$<36.0$ & 37& 38.0& 14&\\
NGC~4278 & 9.2 & 45& 37.9 & 37,46& 40.0& 46&\\
NGC~4472 & 8.9 & 2& 36.7 & 15,37,42 & $<38.8$ & 17&\\
NGC~4494 & 7.7 & 36&$<35.65$& 37& 38.9& 14&\\
NGC~4594 & 9 & 2,7,25& 37.9 & 6,26& 40.2&7&\\
NGC~4621 & 8.4& 8,27& 35.1 & 8& 37.8 &8&\\
NGC~4636 & 8.5&  2& 36.4 & 15,37 & $<38.4$ & 17&\\
NGC~4697 & 8.2& 2,8,29& 35.0& 8& 37.3&8& \\
\hline
\end{tabular}
\vskip 0.3cm Col.(1): Name of the object. Col. (2): Logarithm of
the black hole mass ($\msun$). (4) Logarithm of the nuclear radio
luminosity at 5 GHz ($\ergs$). Col. (6): Logarithm of the X-ray
luminosity in the 2--10 keV band ($\ergs$).

REFERENCES:
(1) Evans et al. 2006; (2) Pellegrini 2005;
(3) Ho et al. 2003; (4) Maoz 2007;
(5) Eracleous et al. 2002; (6) Ho 1999; (7) Pellegrini et al. 2003b;
(8) Wrobel et al. 2008; (9) Wilson \& Yang 2002;
(10) Di Matteo et al. 2003; (11) Pellegrini et al. 2007;
(12) Garcia et al. 2000; (13) Garcia et al. 2005;
(14) Ho et al. 2001; (15) Merloni et al. 2003;
(16) Wu et al. 2007; (17) Loewenstein et al. 2004;
(18) Killeen et al. 1988;
(19) Houghton et al. 2006; (20) Bower et al. 1998;
(21) Verolme et al. 2002; (22) Tonry et al. 2001;
(23) Karachentsev et al. 1996; (24) Nagar et al. 2000;
(25) Kormendy et al. 1996; (26) Hummel et al. 1984;
(27) Tremaine et al. 2002; (28) Ravindranath et al. 2002;
(29) Gebhardt et al. 2003; (30) Pauliny-Toth et al. 1981;
(31) Richstone et al. 2004; (32) Kormendy \& Bender 1999;
(33) Stanek \& Garnavich 1998; (34) Macri et al. 2001;
(35) Crane et al. 1992; (36) Barth et al. 2002;
(37) Nagar et al. 2002; (38) Capetti et al. 2002;
(39) Donato et al. 2004; (40) Giovannini et al. 2001;
(41) Katz-Stone \& Rudnick 1997; (42) Ho \& Ulvestad 2001;
(43) Ho 2002; (44) Terashima et al. 2002; (45) Magorrian et al. 1998;
(46) Terashima \& Wilson 2003.

\end{table*}

\clearpage

\begin{deluxetable}{lllllllllllllll}
\rotate
\tablewidth{23cm}
\tablecaption{Modeling of Individual Objects}
\tabletypesize{\scriptsize}
\tablehead{Object &$M (\msun)$ &$L_{\rm 2-10keV} (L_{\rm Edd})$ &$L_{\rm X,crit} (L_{\rm Edd})$
&Ref. &$\dot{m}_{\rm out}$ &$R_{\rm out} (r_g)$ &$\dot{m}_{\rm jet}$ &$\Gamma$
&$\theta$ ($^{\circ}$)&$\epsilon_{\rm e}$ &$\epsilon_{\rm B}$ &$p$ &Origin of & Consistent \\&&&&&&&&&&&&&X-rays&with YC05? \\
(1) &
(2) &
(3) &
(4) &
(5) &
(6) &
(7) &
(8) &
(9) &
(10) &
(11) &
(12) &
(13) &
&
}
\startdata
IC~1459      &$2\times10^9$    &$1.7\times 10^{-7}$ &$1.2\times10^{-7}$  &19,20 &$2\times10^{-4}$&$2\times10^4$ &$7\times10^{-7}$    &10 &30 &0.028 &0.02 &1.9 &jet  &no\\
IC~4296       &$1\times10^9$   &$1.3\times 10^{-6}$ &$1.3\times 10^{-7}$ &1 &$1.1\times10^{-3}$ &$2\times 10^4$ &$7\times10^{-6}$ &10 &60 &0.2&0.02 &2.1 &ADAF &yes\\
M~32         &$2.5\times 10^6$ &$3  \times 10^{-9}$ &$3.6\times 10^{-7}$ &4 &$4.5\times10^{-4}$ &$2\times10^5$  &$1\times10^{-8}$      &10 &40 &0.28&0.01   &2.5 &jet  &yes\\
M~81         &$7\times10^7$    &$2.3\times10^{-6}$  &$2\times10^{-7}$    &3,13,14,21 &$2\times10^{-3}$ &$2\times10^4$ &$2.4\times10^{-7}$ &10&25 &0.25 &0.02&1.8 &jet &no\\
M~84          &$1.6\times 10^9$ &$1.1\times 10^{-8}$ &$1.2\times 10^{-7}$ &2 &$2.5\times10^{-4}$ &$4\times10^4$  &$8\times10^{-7}$ &10 &63 &0.2&0.02  &2.35 &jet  &yes\\
M~87         &$3.4\times10^9$  &$8.1\times10^{-8}$  &$1\times10^{-7}$           &15,16,3 &$1\times10^{-3}$ &$1\times10^5$            &$1\times10^{-8}$           &20 &19 &0.14 &0.02 &2.5 &jet &yes\\
NGC~315      &$3.1\times10^8$  &$1.5\times 10^{-5}$ &$1.6\times 10^{-7}$ &5,6 &$4\times10^{-3}$ &$2\times10^4$  &$7\times10^{-6}$      &10 &38 &0.15&0.02    &2.5 &ADAF &yes\\
NGC~1052     &$1.26\times10^8$ &$5.9\times 10^{-6}$ &$1.85\times10^{-7}$ &3 &$1.5\times10^{-3}$ &$1\times10^4$$^a$        &$8\times10^{-5}$ &10 &60 &0.2&0.02    &2.3 &ADAF &yes\\
NGC~3998     &$7\times10^8$    &$3\times10^{-6}$    &$1.4\times10^{-7}$  &3,7,22 &$6\times10^{-4}$ &$6\times10^2$          &$9.8\times10^{-8}$ &10 &20 &0.2 &0.02 &1.8 &jet  &no\\
NGC~4203     &$1\times10^7$           &$1.85\times10^{-5}$ &$2.84\times10^{-7}$ &8,3 &$6\times10^{-3}$&$1\times10^3$        &$1.5\times10^{-6}$ &10 &25 &0.1 &0.02 &2.2 &ADAF &yes\\
NGC~4261     &$4.9\times10^9$   &$4\times10^{-6}$    &$1.5\times10^{-7}$ &17 &$1.7\times10^{-3}$ &$2\times10^4$ &$5\times10^{-6}$    &10 &63 &0.3  &0.01 &2.45 &ADAF &yes\\
NGC~4579\tablenotemark{a} &$4\times10^6$    &$3\times10^{-4}$    &$3.3\times10^{-7}$  &9,8 &$2.3\times10^{-2}$        &$2\times10^2    $   &$8\times10^{-5}$  &10 &60 &0.2&0.02 &2.3  &ADAF &yes\\
NGC~4594     &$1\times10^9$           &$1.2\times10^{-7}$  &$1.3\times10^{-7}$  &10,11,9 &$5\times10^{-4}$ &$2\times10^4$ &$4\times10^{-7}$ &10&45 &0.3 &0.02&1.8  &jet &yes\\
NGC~4621     &$2.7\times10^8$  &$1.9\times10^{-9}$  &$1.6\times10^{-7}$  &12  &$6\times10^{-6}$ &$2\times10^4$  &$4\times10^{-8}$  &10 &60 &0.5 &0.02&2.3  &jet &yes\\
NGC~4697     &$1.7\times10^8$  &$1\times10^{-9}$           &$1.76\times10^{-7}$ &12 &$6\times10^{-6}$ &$2\times10^4$   &$3.5\times10^{-8}$ &10 &60 &0.3&0.02&2.1 &jet &yes\\
NGC~6251     &$6\times 10^8$   &$5\times10^{-5}$    &$1.4\times10^{-7}$  &18 &$1.8\times10^{-2}$ &$2\times10^4$ &$4\times10^{-6}$    &10 &30 &0.2&0.02  &2.4  &ADAF &yes\\

\enddata
\tablecomments{
Col. (1): Name of object.
Col. (2): Mass of the black hole.
Col. (3): Luminosity in the 2--10 keV band.
Col. (4): Critical luminosity from eq. (\ref{eq:critlum}).
Col. (5): References.
Col. (6): Mass accretion rate at $R_{\rm out}$.
Col. (7): $R_{\rm out}$.
Col. (8): Mass lost rate in the jet.
Col. (9): Lorentz factor of the jet.
Col. (10): Viewing angle of the jet.
Col. (11): Energy density of accelerated electrons.
Col. (12): Energy density of amplified magnetic field.
Col. (13): Spectral index of energy distribution of electrons.
}
\tablenotetext{a}{For this source, $s=0.1$, i.e $\dot{M}=\dot{M}_{\rm out}(R/R_{\rm out})^{0.1}$}
\tablerefs{
(1) Pellegrini et al. 2003a;
(2) Pellegrini et al. 2005;
(3) Maoz 2007;
(4) Ho et al. 2003;
(5) Worrall et al. 2007;
(6) Gu et al. 2007;
(7) Ptak et al. 2004;
(8) Terashima et al. 2002;
(9) Ho 1999;
(10) Hummel et al. 1984;
(11) Pellegrini et al. 2003b;
(12) Wrobel et al. 2008;
(13) Sch\"odel et al. 2007;
(14) Markoff et al. 2008;
(15) Wilson \& Yang 2002;
(16) Evans et al. 2006;
(17) Zezas et al. 2005;
(18) Evans et al. 2005;
(19) Fabbiano et al. 2003;
(20) Trinchieri \& McDowell 2003;
(21) Pellegrini et al. 2000a;
(22) Pellegrini et al. 2000b.
}
\end{deluxetable}


{}

\clearpage

\begin{figure*}
\epsscale{0.35} \plotone{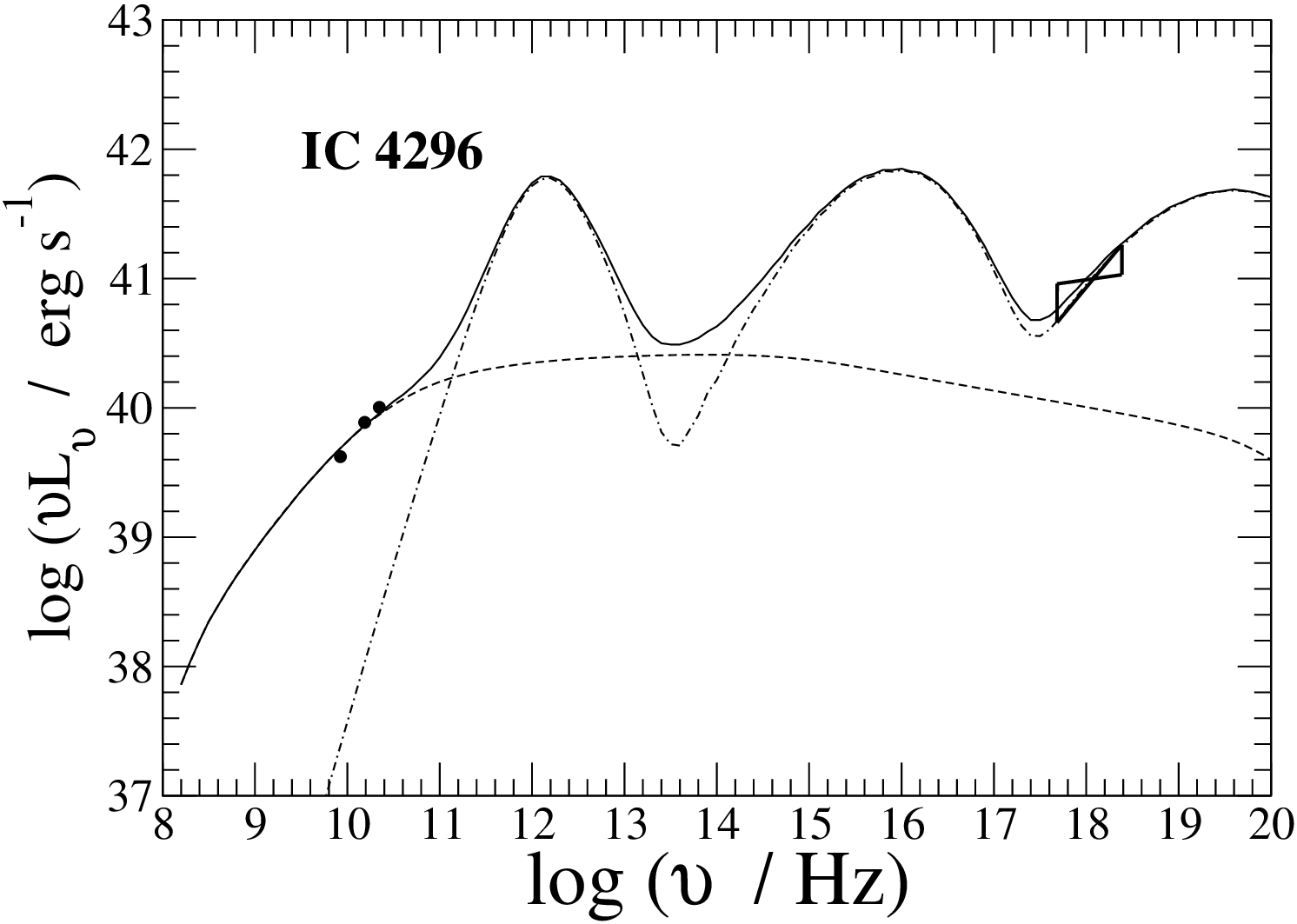}\hspace{1cm} \epsscale{0.35}
\plotone{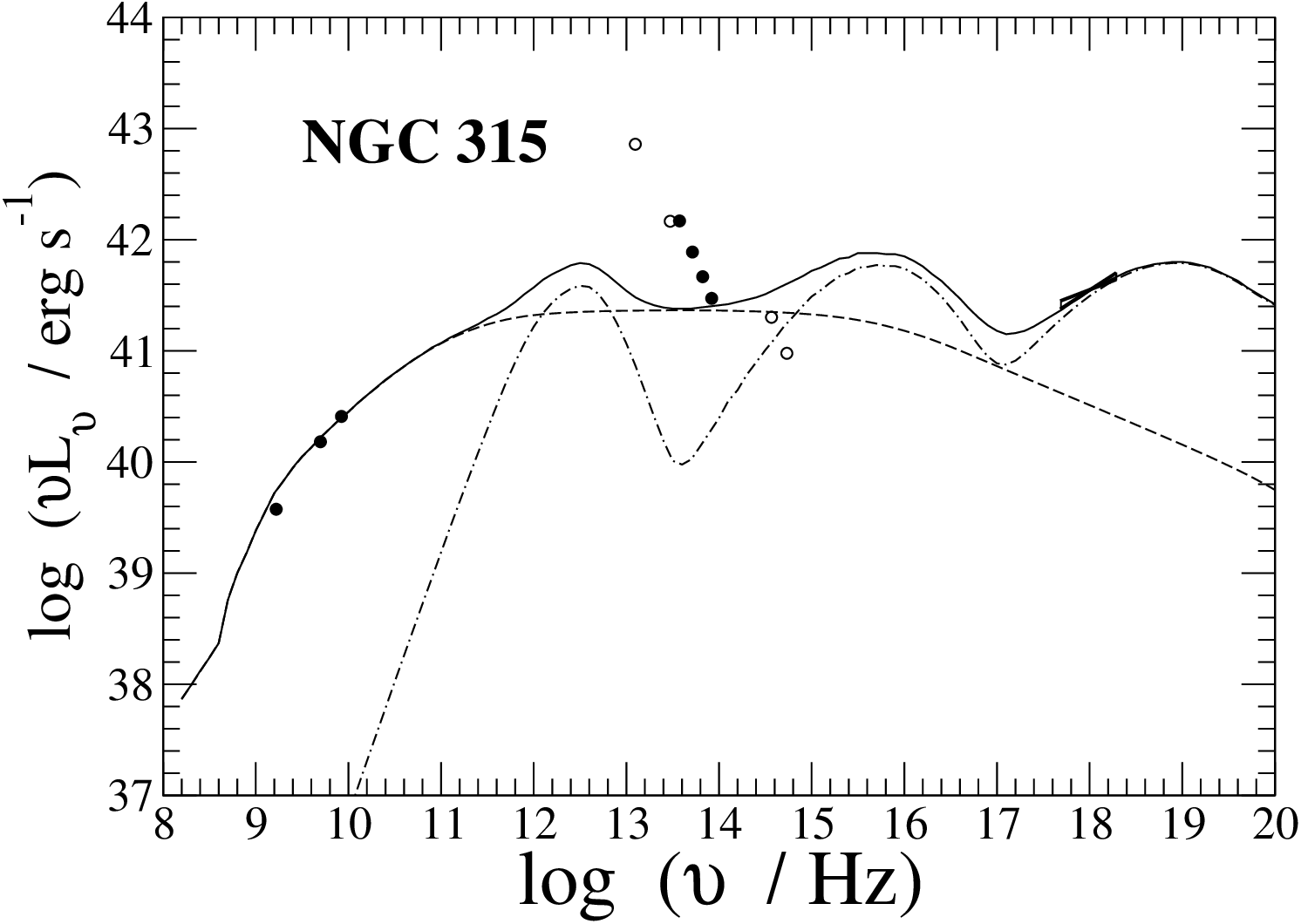} \vspace{0.2in} \epsscale{0.35}
\plotone{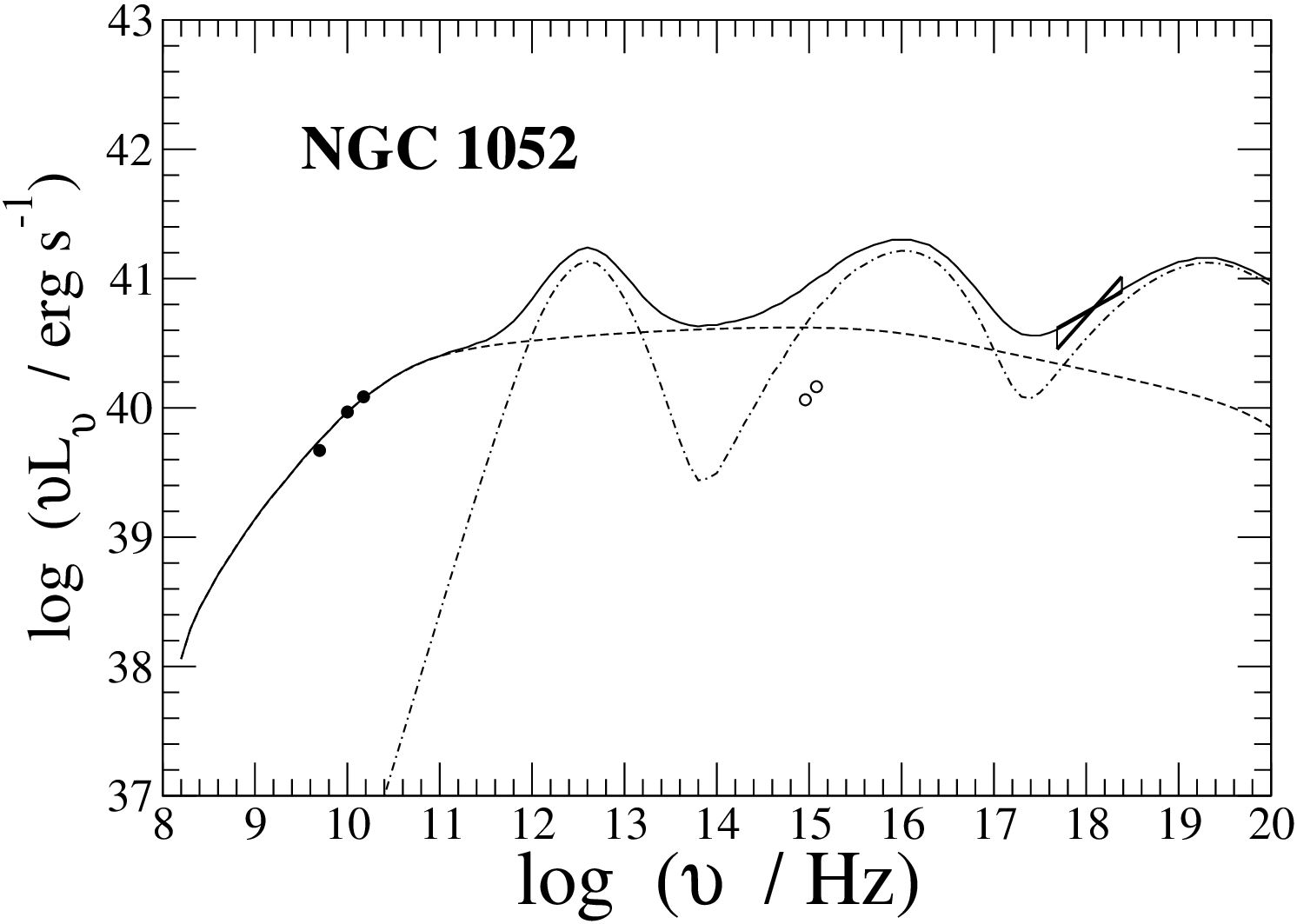}\hspace{1cm} \epsscale{0.35}
\plotone{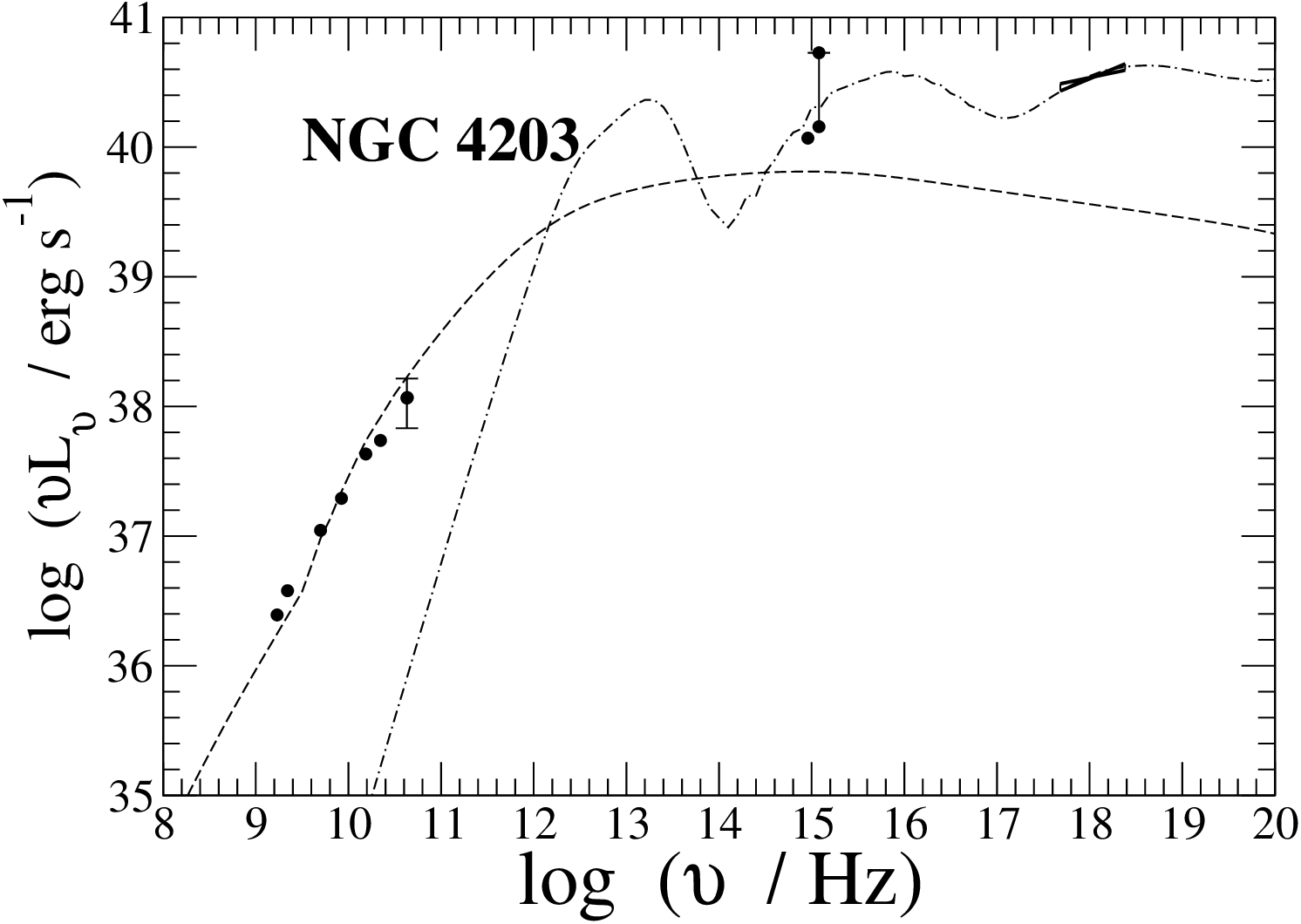}\vspace{0.2in} \epsscale{0.35}
\plotone{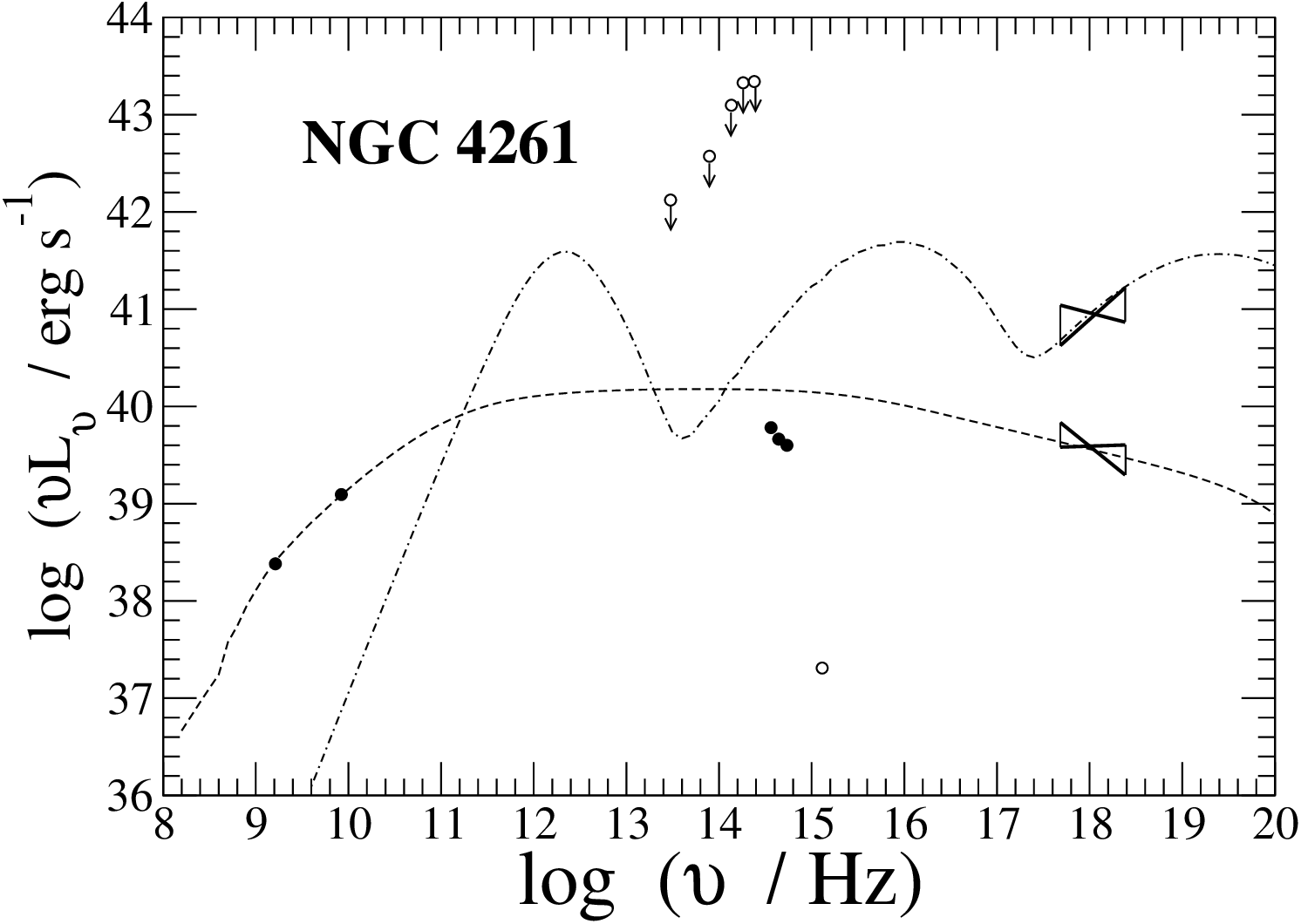}\hspace{1.0cm} \epsscale{0.35} \plotone{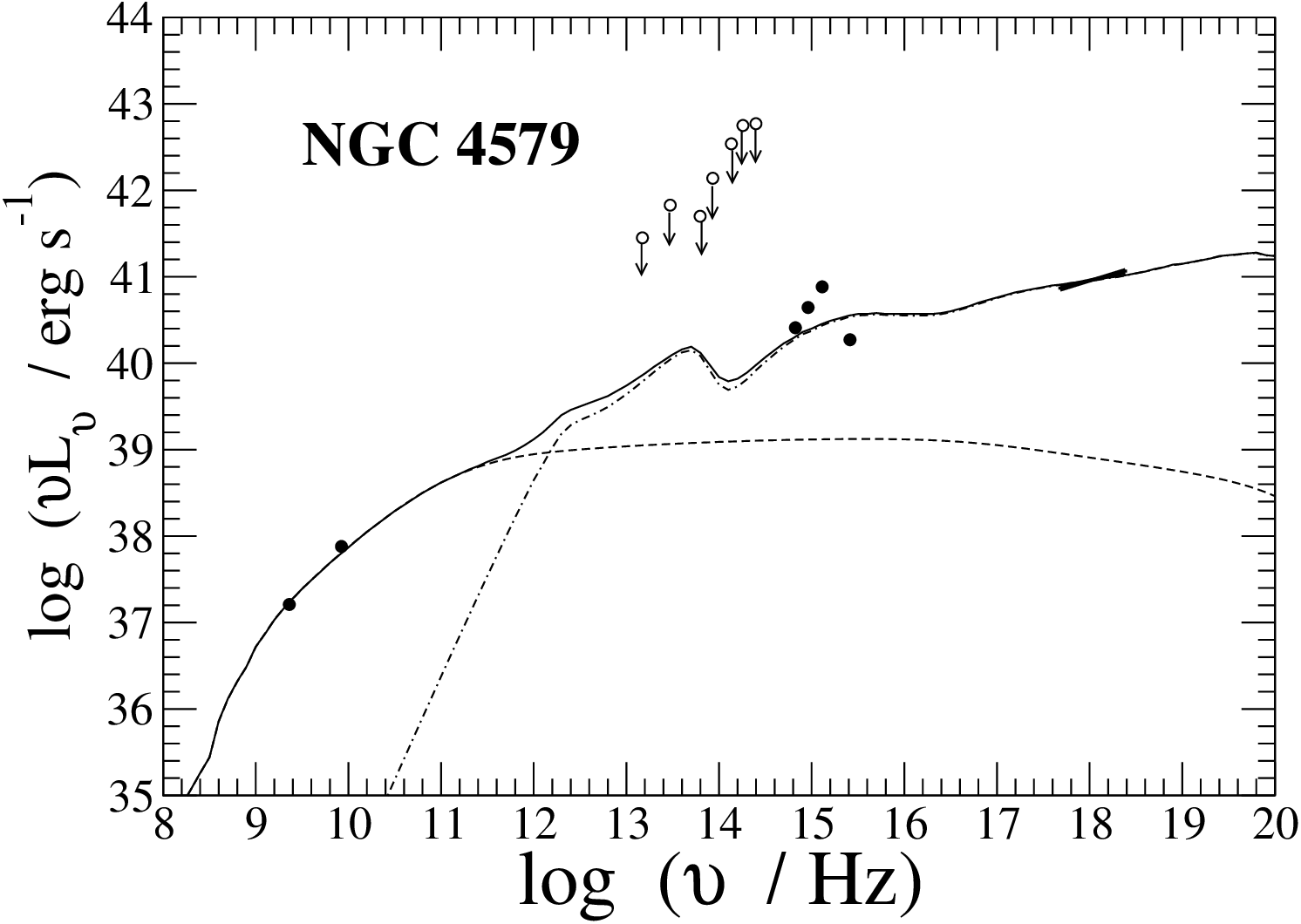}
\vspace{0.3in} \epsscale{0.35} \plotone{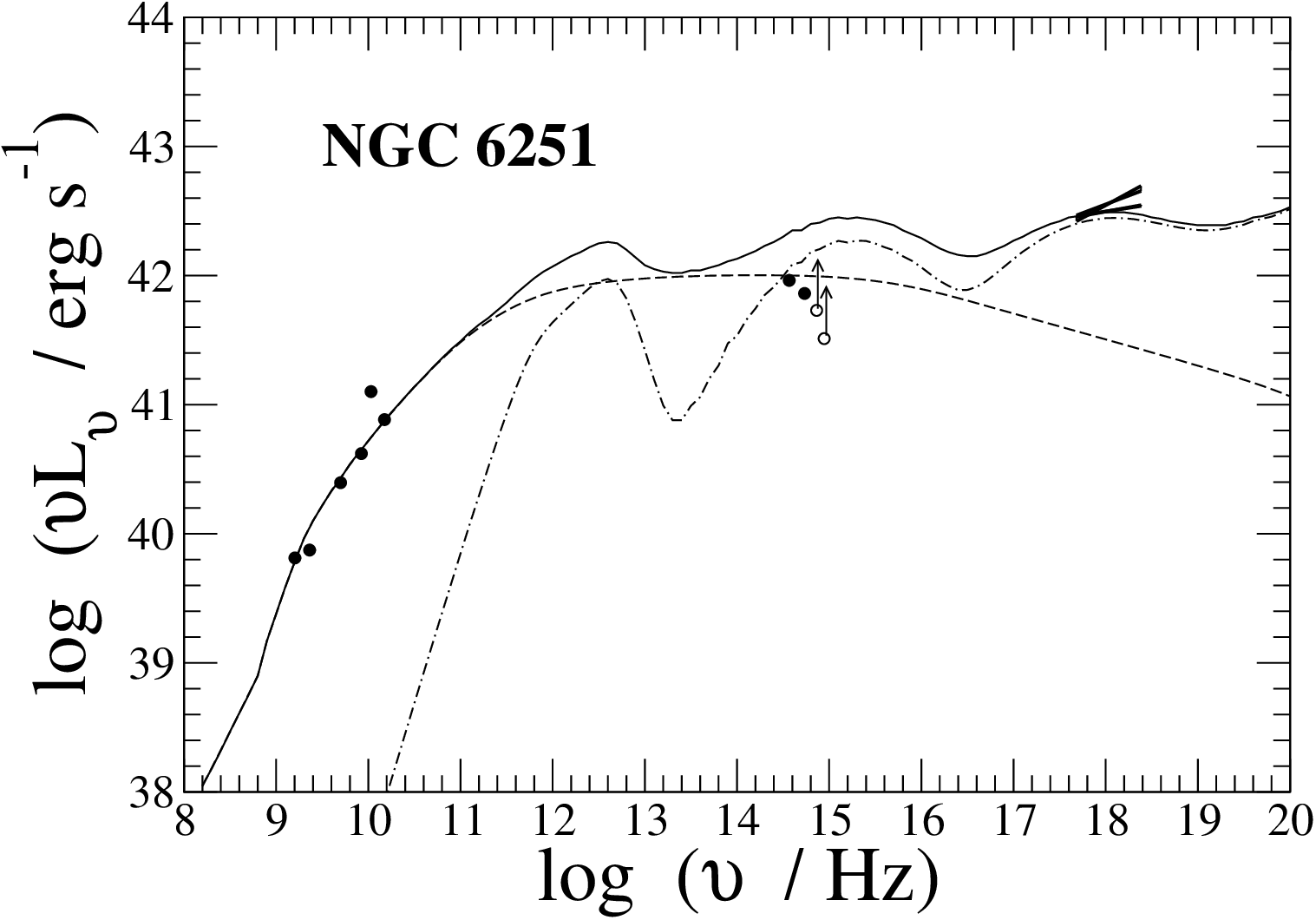} \caption{The
ADAF-dominated sources. The dot-dashed lines show the emitted
spectra of ADAFs, the dashed lines show the spectra of jets, and the
solid lines show their sum. Reliable data points are plotted as
solid symbols, whereas points severely affected by host galaxy
contamination or extinction are plotted as open symbols.  All of
these sources are consistent with the prediction of YC05 (see Table
2). }
\end{figure*}

\begin{figure*}
\epsscale{0.32} \plotone{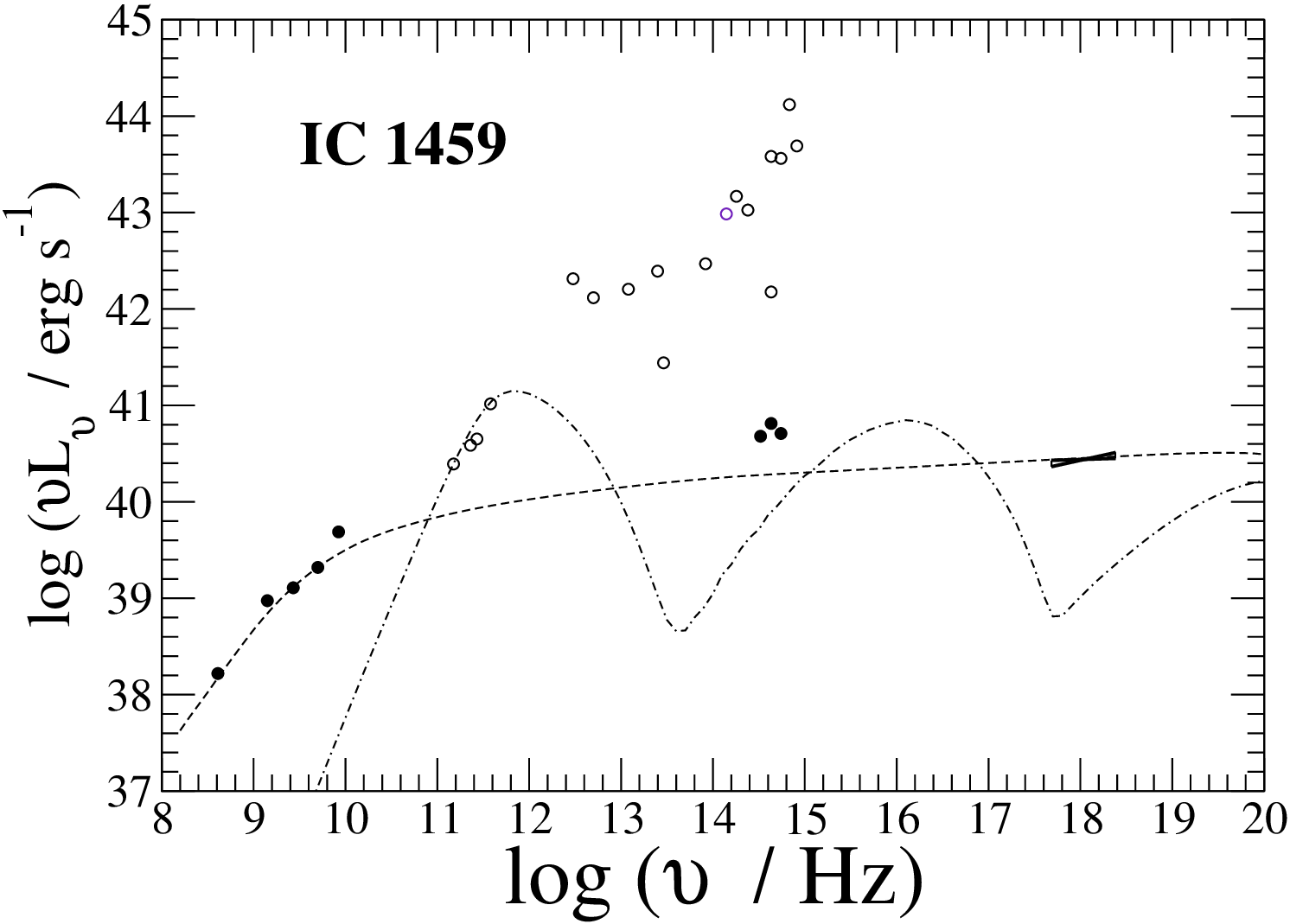}\hspace{1cm} \epsscale{0.32}
\plotone{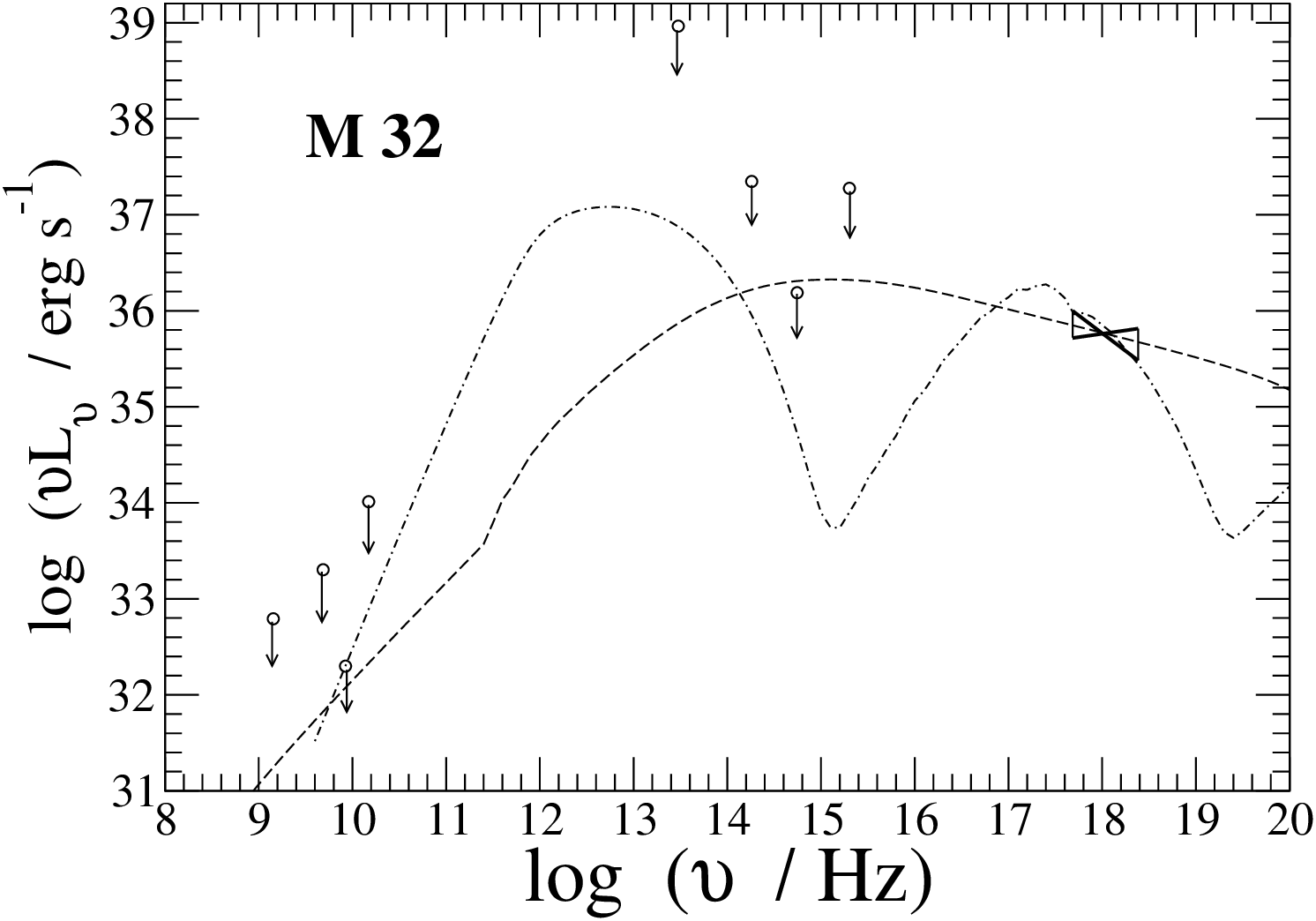} \vspace{0.2in} \epsscale{0.32} \plotone{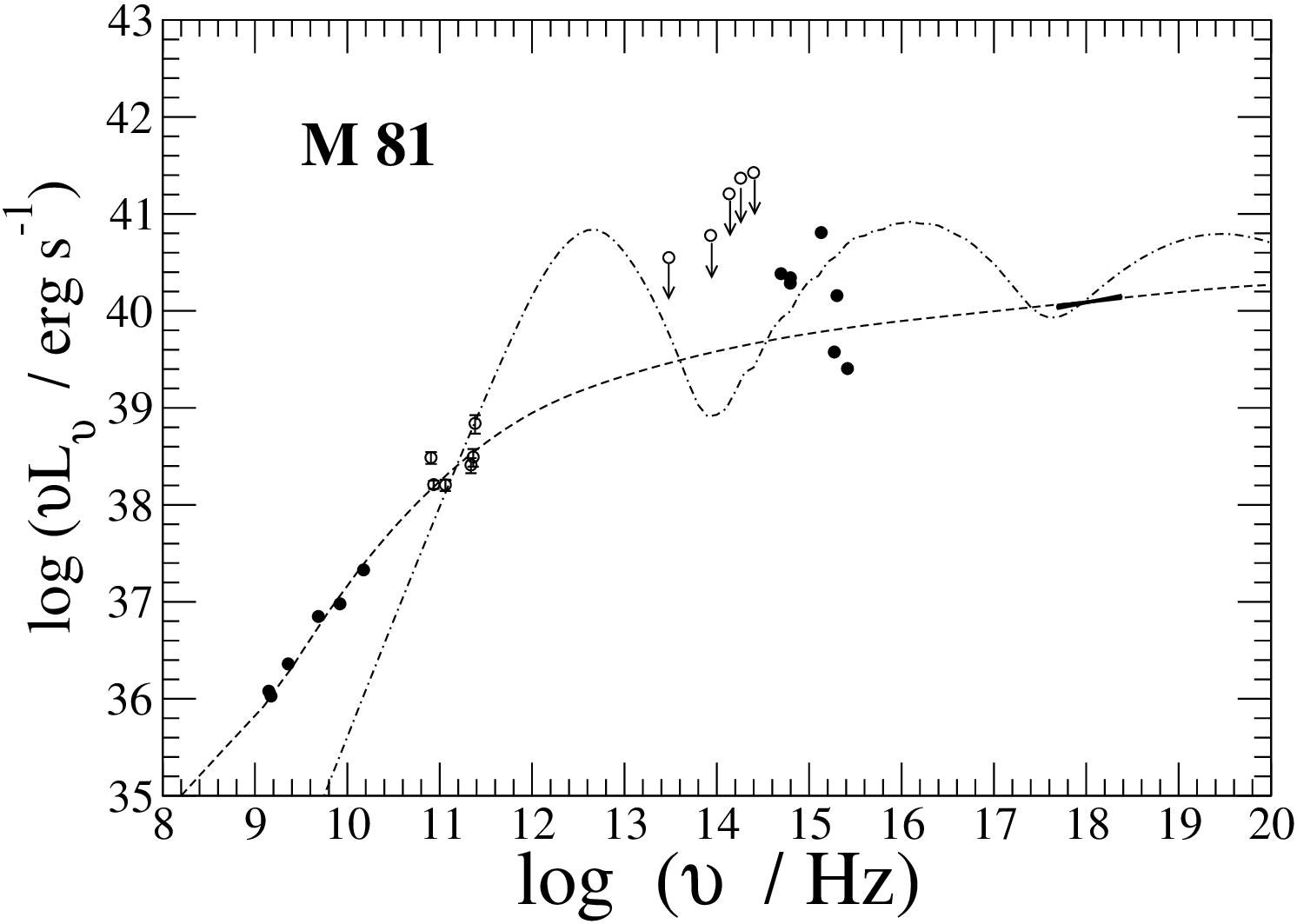}
\hspace{1cm} \epsscale{0.32} \plotone{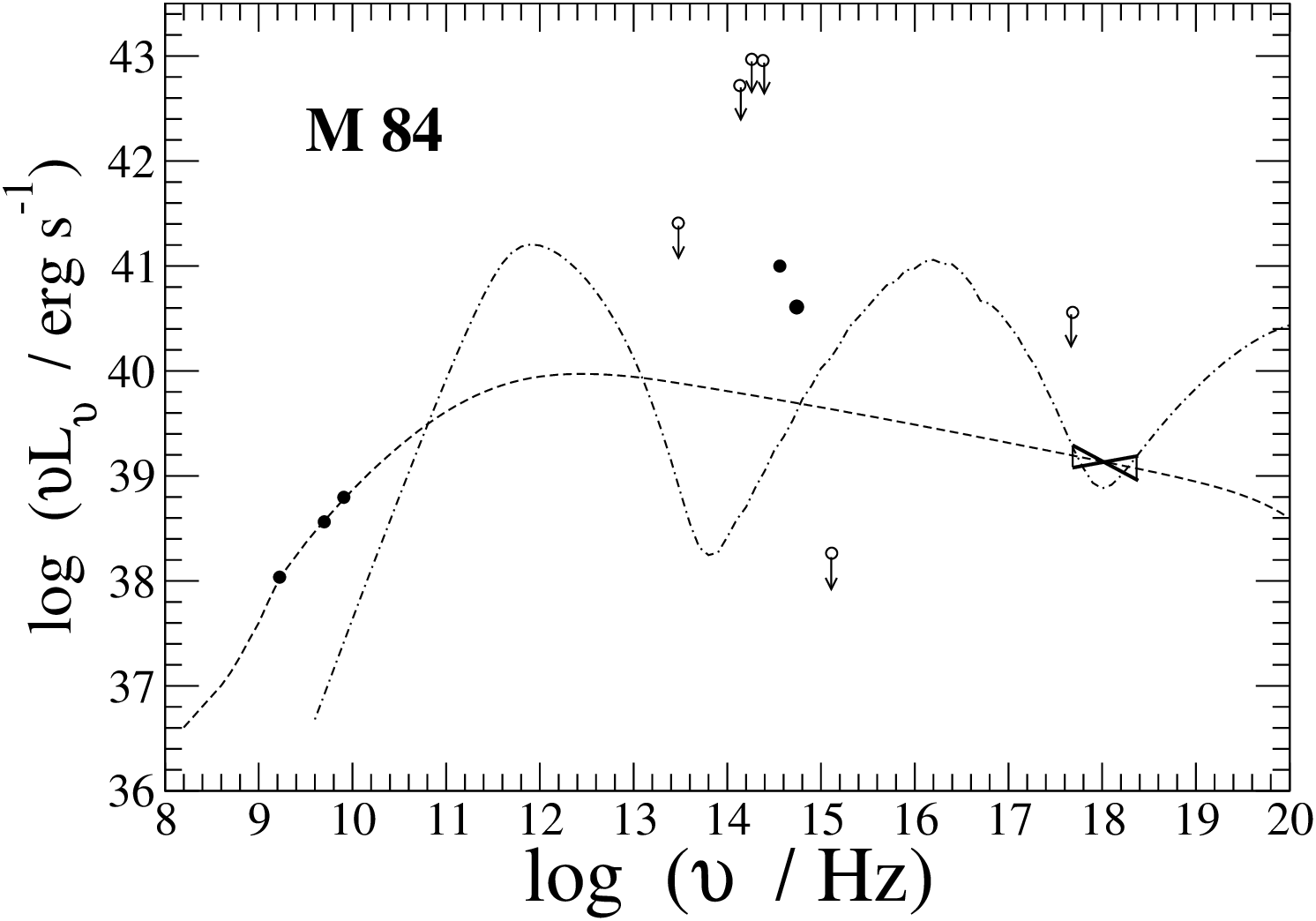} \vspace{0.2in}
\epsscale{0.32} \plotone{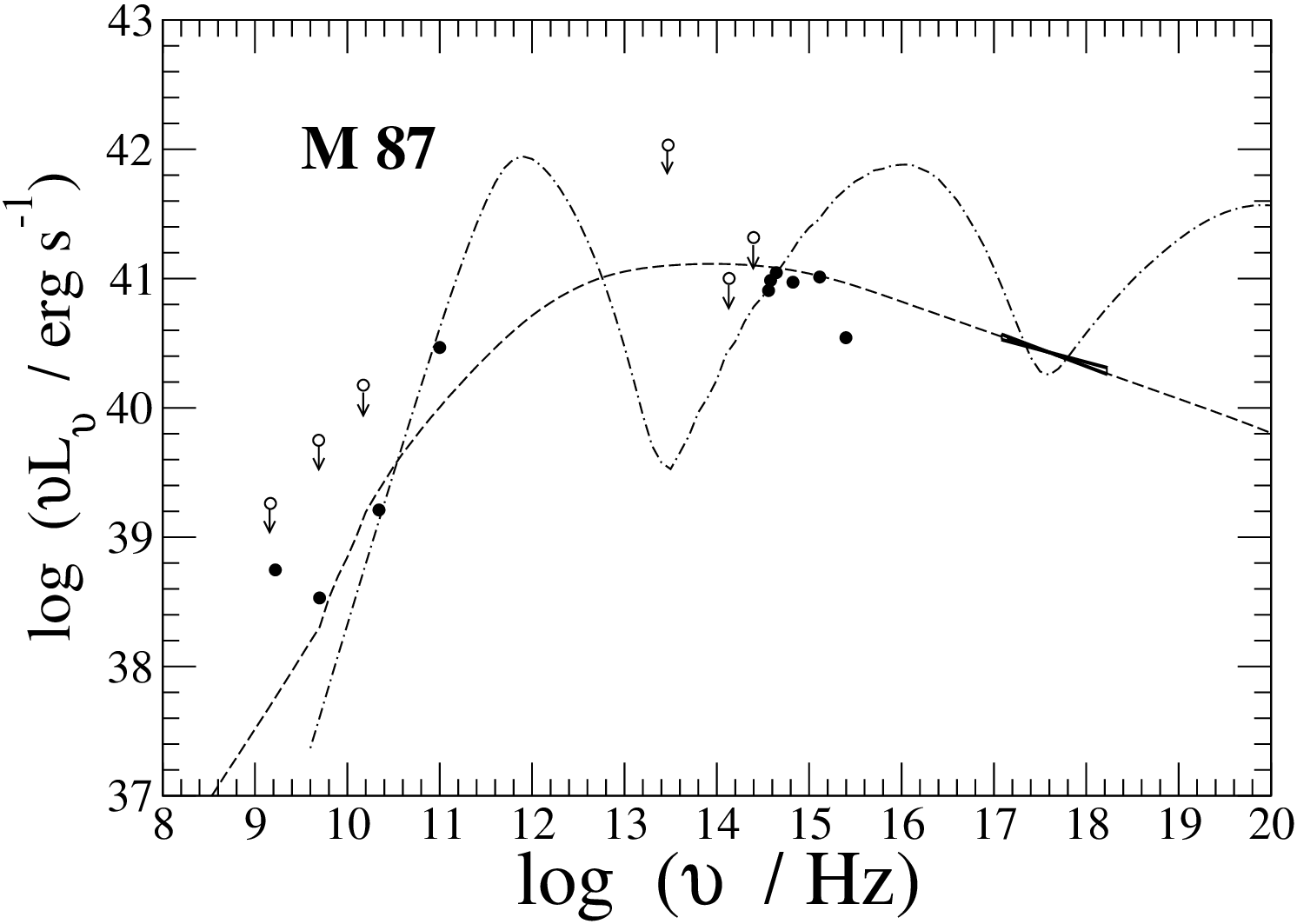}\hspace{1cm} \epsscale{0.32}
\plotone{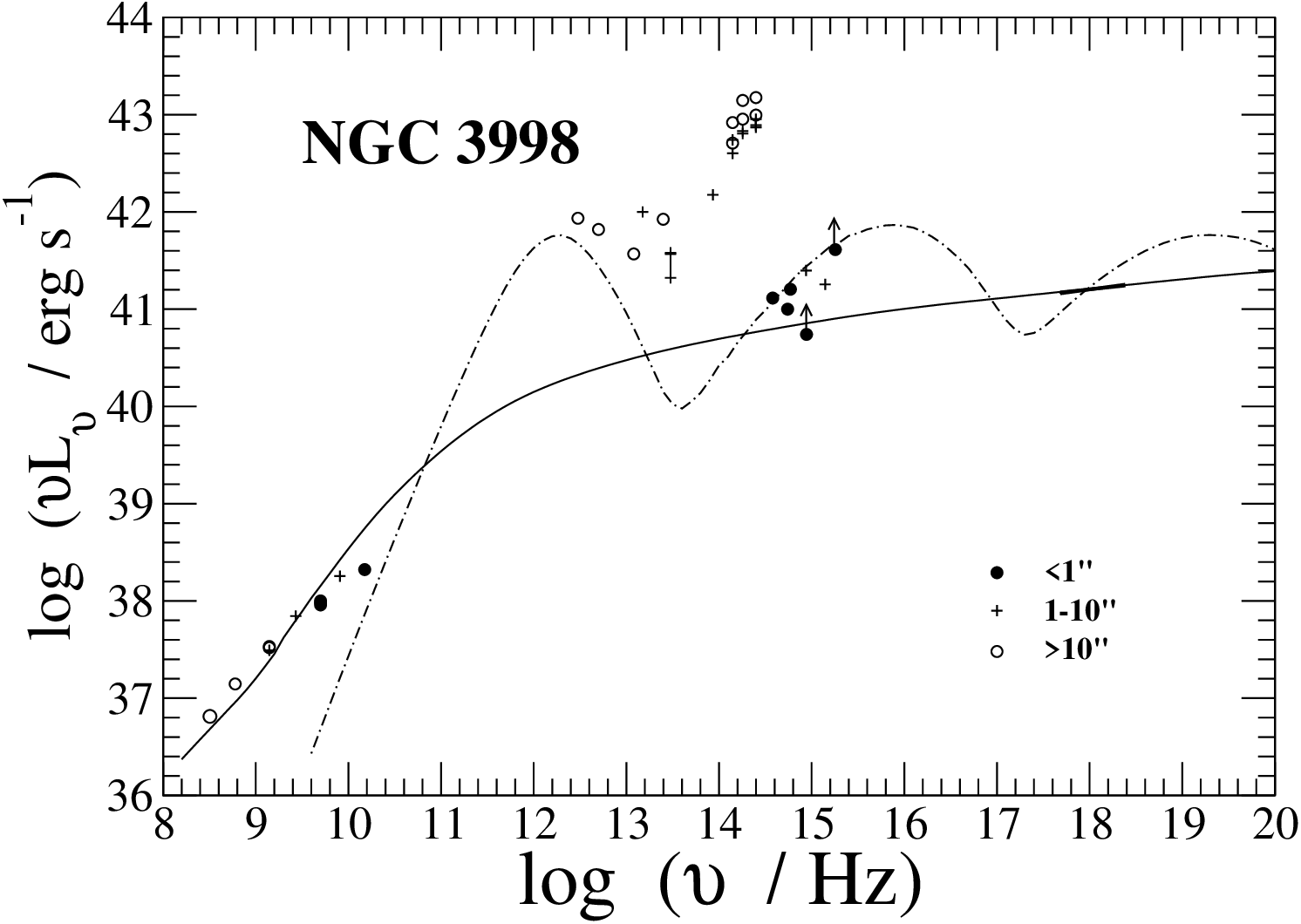} \vspace{0.2in} \epsscale{0.32} \plotone{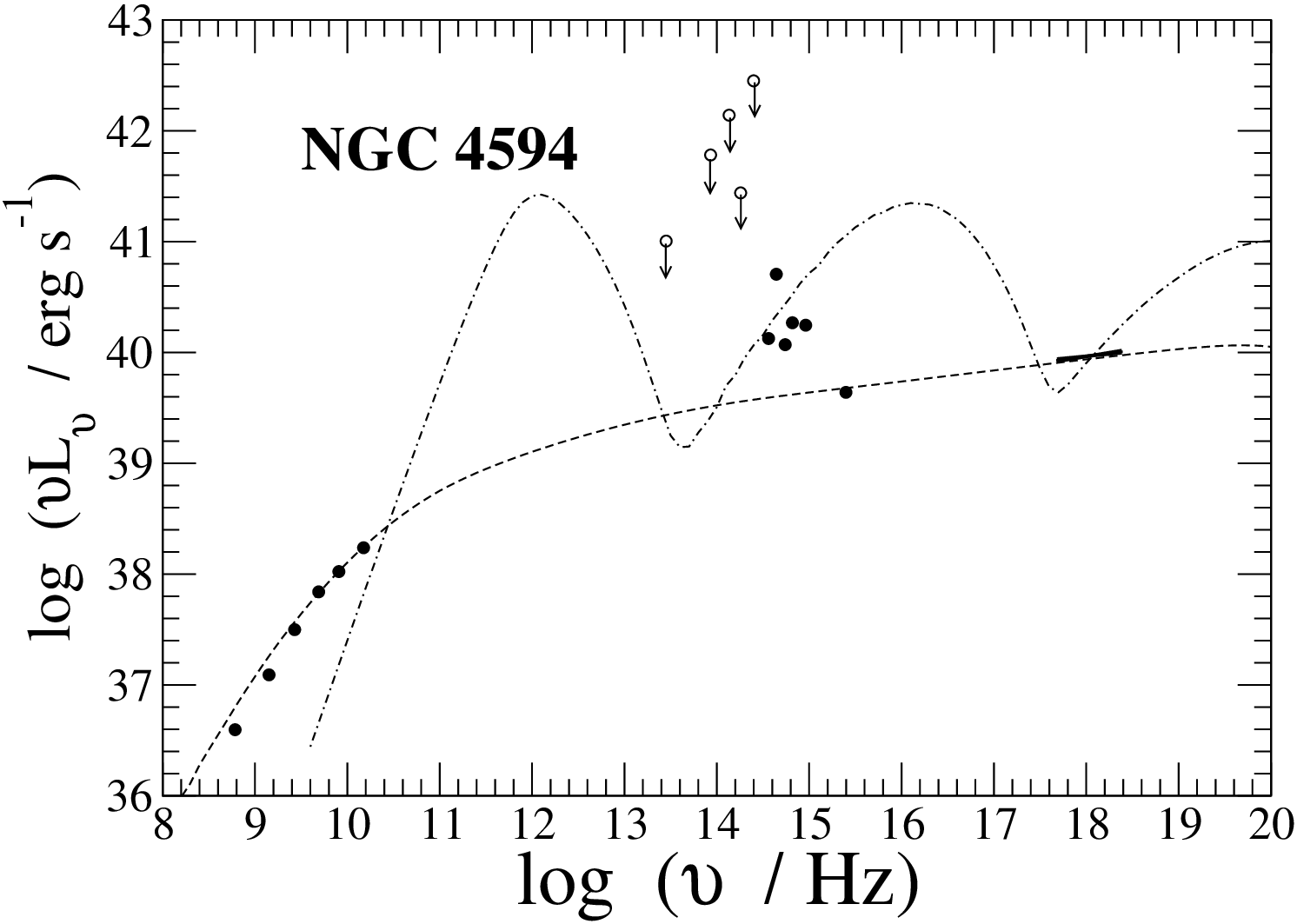}
\hspace{1cm} \epsscale{0.32} \plotone{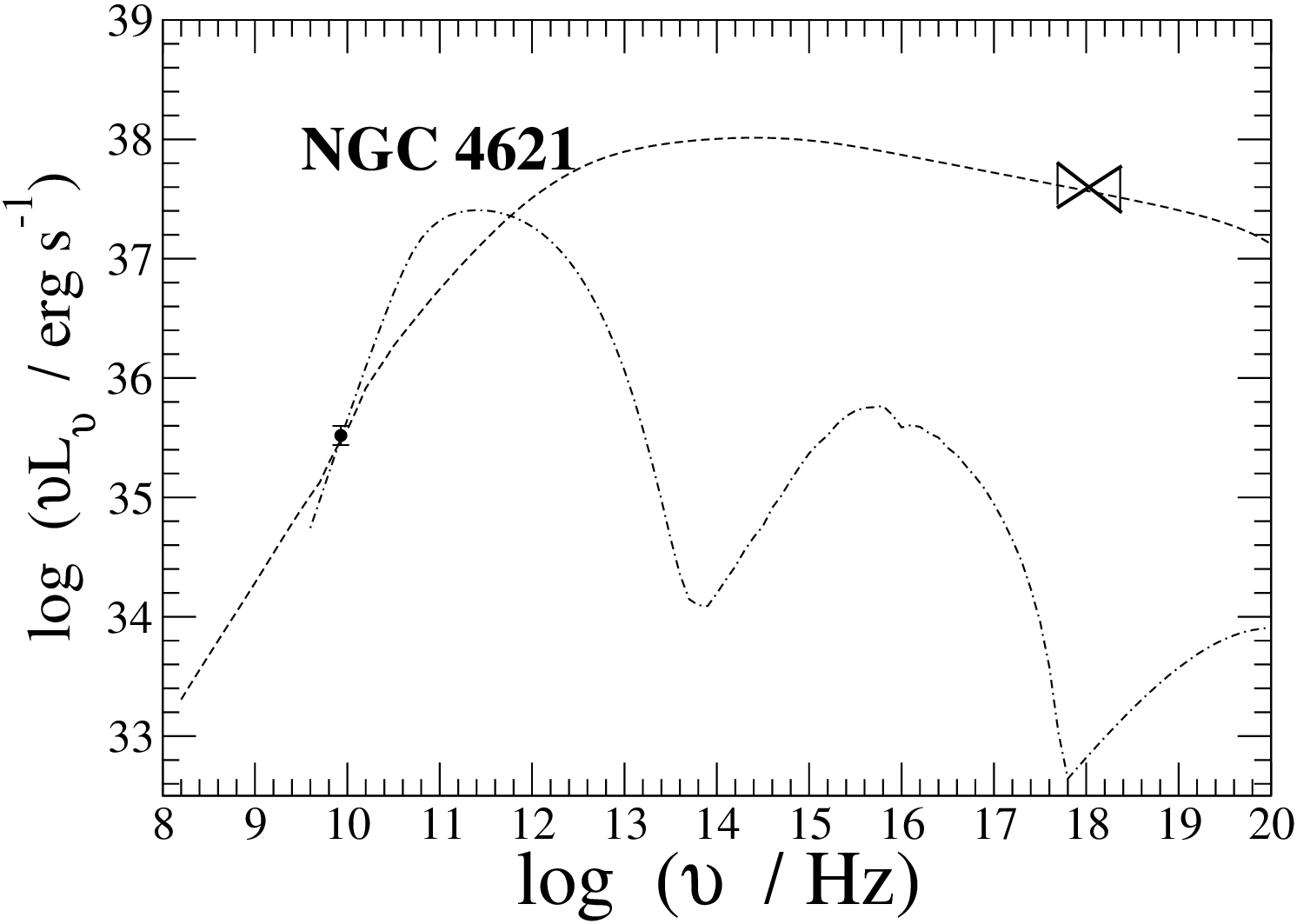}\vspace{0.2in}
\epsscale{0.32} \plotone{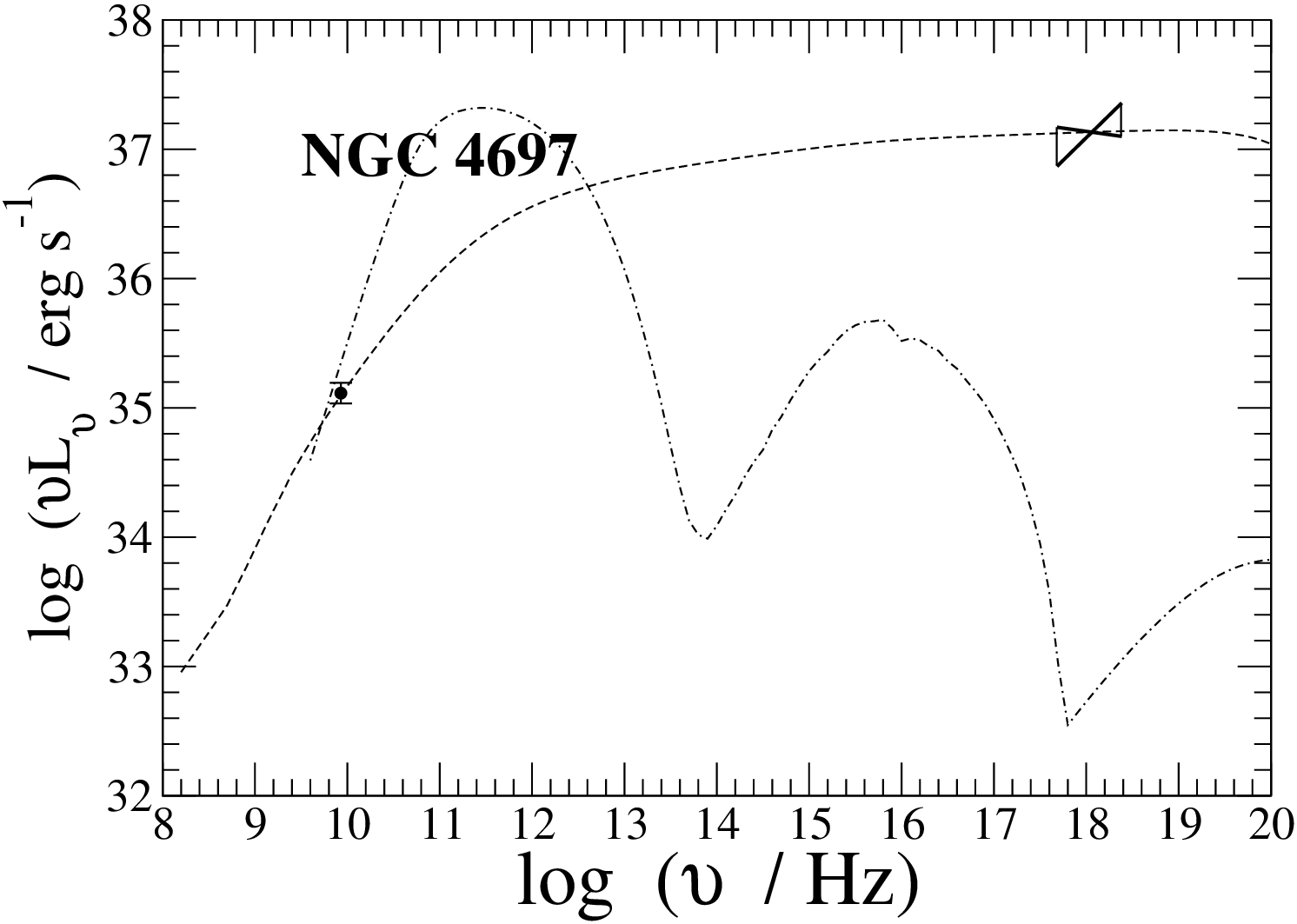} \caption{The jet-dominated
sources. The dot-dashed lines show the emitted spectra of ADAFs, and
the dashed lines show the spectra of jets. Reliable data points are
plotted as solid symbols, whereas points severely affected by host
galaxy contamination or extinction are plotted as open symbols. All
of these sources, with the exception of IC~1459, M~81, and NGC~3998,
are consistent with the prediction of YC05 (see Table 2). }
\end{figure*}

\end{document}